\documentclass[12pt,preprint]{aastex}

\newcommand{\et}{et al.}

\newcommand{\ha}{H$\alpha$}

\newcommand{\rhalf}{R$_{0.5}$}

\newcommand{\solar}{\ifmmode_{\sun}\;\else$_{\sun}\;$\fi}

\newcommand{\HII}{H$\,${\sc ii}}

\begin{document}

\title{Compact Star Clusters in Nearby Dwarf Irregular Galaxies\footnote{\rm Based
on observations with
the NASA/ESA  Hubble Space Telescope, obtained at the Space Telescope
Science Institute, which is  operated  by AURA, Inc.,
under NASA contract NAS5-26555.}
}

\author{Olivia H. Billett\footnote{\rm Current address:  236 Park Street, New Haven, CT
06511 (olivia.billett@yale.edu)}, \ Deidre A. Hunter}
\affil{Lowell Observatory, 1400 West Mars Hill Road, Flagstaff, Arizona 86001
USA}
\email{olivia@lowell.edu, dah@lowell.edu}

\and

\author{Bruce G.\ Elmegreen}
\affil{IBM T.\ J.\ Watson Research Center, PO Box 218, Yorktown Heights,
New York 10598 USA}
\email{bge@watson.ibm.com}

\begin{abstract}

Nearby dwarf irregular galaxies were searched for compact star
clusters using data from the {\it HST} archives. Eight of the 22
galaxies in our sample were found to host compact clusters of some
type. Three of these have populous clusters, with M$_V<-9.5$ at a fiducial
age of 10 Myr, and the same three also have super-star clusters,
with M$_V<-10.5$ at 10 Myr. Four other dwarf galaxies, two of
which contain populous and super-star clusters, are also
considered using data in the literature.  The results suggest that
galaxies fainter than M$_B=-16$ or with star formation rates less
than 0.003 M\solar\ yr$^{-1}$ kpc$^{-2}$ do not form populous or
super-star clusters, and that even the brighter and more active
dwarfs rarely form them. Yet when they do form, the associated
star formation activity is very high, with numerous compact
clusters of similar age in the same complex and evidence for a
galaxy-wide perturbation as the trigger.  This tendency to
concentrate star formation in localized regions of high column
density is consistent with previous suggestions that self-gravity
must be strong and the pressure must be high to allow a cool phase
of gas to exist in equilibrium.

Statistical considerations emphasize the peculiarity of super star
clusters in dwarf galaxies, which are too small to sample the
cluster mass function to that extreme.  We suggest that triggered
large-scale flows and ambient gravitational instabilities in the
absence of shear make the clouds that form super-star clusters in
small galaxies. This is unlike the case in spiral galaxies where
density wave flows and scale-free compression from turbulence seem
to dominate. Further comparisons with spiral galaxies give insight
into Larsen \& Richtler's relation between the star formation rate
per unit area and the fraction of young stars in massive dense
clusters.  We suggest that this relation is the result of a
physical connection between maximum cluster mass, interstellar
pressure, interstellar column density, and star formation rate,
combined with a size-of-sample effect.

\end{abstract}


\section{Introduction}

The properties of super-star clusters set them apart.  They are extremely
compact and luminous clusters, with full-width at half maximum (FWHM)
of less than 15 pc and
M$_V$ brighter than about $-$10 (see, for example,
van den Bergh 1971; Arp \& Sandage 1985; Melnick, Moles, \& Terlevich 1985;
Holtzman \et\ 1992; Whitmore \et\ 1993).
If they are bound and contain low mass stars, they resemble what we would expect a
globular cluster to be like when it was young. R136, for example,
the star cluster in the 30 Doradus \HII\ complex in the LMC, is
the nearest example of a small super-star cluster. It has
a half-light radius \rhalf\ of 1.7 pc, an M$_V$ at an age of
2 Myr of $-$11, and a mass of $6\times10^4$ M\solar\ if
the mass spectrum continues down to 0.1 M\solar\ stars
(Hunter \et\ 1995).
Therefore, the super-star clusters are an extreme mode of star formation.

In spite of the fact that the Milky Way has not been able to form
a cluster as compact and luminous as a globular cluster for some
10 Gyr (although there may be one forming now---Kn\"odlseder
2000), young and old super-star clusters have been found in tiny
irregular galaxies. Six super-star clusters are known in five
nearby dwarf irregular galaxies and are inferred to be present,
though still embedded, in 4 others. NGC 1569, for example, with an
integrated M$_V$ of only $-18$, nevertheless, hosts two young
super-star clusters with individual M$_V$ of $-$14 and $-$13 (Arp
\& Sandage 1985; O'Connell, Gallagher, \& Hunter 1994). WLM, with
a galactic M$_V$ of only $-$14, contains a bonafide globular
cluster that is 14.8 Gyr old (Hodge \et\ 1999). So, what are the
conditions that allow small irregular galaxies to form super-star
clusters when the Milky Way cannot? Do super-star clusters require
the same extraordinary circumstances to develop as globular
clusters did, and what are those conditions?

Populous clusters,
less extreme cousins to the super-star clusters, have also been found in large numbers
in the Magellanic Clouds (see, for example, Gascoigne \& Kron 1952;
Hodge 1960, 1961; Bica \et\ 1996).
These objects are less luminous and less massive than the
super-star clusters but more extreme in luminosity and compactness than open clusters.
A young populous cluster is exemplified by
NGC 1818 in the LMC. At an age of about 20 Myr, it has an M$_V$ of
$-$9.3, an \rhalf\ of 3.2 pc, and a mass of 3$\times10^4$ M\solar\ (Hunter \et\ 1997).
Besides the Magellanic Clouds, 14 populous clusters
have been identified in the irregular NGC 1569 (Hunter \et\ 2000),
but none have been found in galaxies like IC 1613 (Hodge 1978)
and NGC 6822 (Hodge 1977). (The star clusters identified in IC 1613 and
NGC 6822 by Hodge are open clusters although one in NGC 6822 has a globular-like
metallicity [Cohen \& Blakeslee 1998]).
So, what conditions are necessary for the formation of populous clusters?

The high resolution of Hubble Space Telescope {\it (HST)} is essential
to surveying even nearby galaxies for compact clusters. Since a
fundamental characteristic of these clusters is their compactness, it is
easy to mistake them for a single star with inadequate resolution. The
closest super-star cluster, R136 in the LMC, has a half-light diameter of
about 13\arcsec.  At a distance of 0.7 Mpc, the lower
end of the range studied here, R136 would have a radius of 1\arcsec, which
pushes the limit of typical ground-based telescopes. {\it HST}'s factor of 10
higher resolution allows us to detect and measure clusters up to 10
times more distant. The problem with resolution is exemplified by the
disagreements based on ground-based data
over whether the core of R136 was a super massive 2000 M\solar\ star
or a cluster of very massive stars (Cassinelli, Mathis, \& Savage 1981)
and whether the two super-star clusters in NGC 1569 were really
star clusters or bright single stars in that galaxy (Arp \& Sandage 1985).
In both cases the objects are clusters of stars rather than single stars,
but studies using ground-based data could not definitely determine this.
Due to these inadequacies, a survey
for compact clusters beyond the Magellanic Clouds benefits greatly from
{\it HST}.

We have used images from the {\it HST} archives of dwarf irregular
galaxies within 7 Mpc to study cluster-producing environments.
Unfortunately, since we are using archival data, we did not have a
choice in the pointing of the telescope, the filters used, or the
exposure times, and the characteristics of what was available
placed limits on our search. Promising galaxies that were imaged
for a teasing 0.5 s had to be passed over, as did those with only
one filter. Only nearby dwarfs galaxies that could be reasonably
searched were included in the survey.

\section{Data}

The data used were taken with the {\it HST} and consist of Wide
Field and Planetary Camera 2 (WFPC2) images of 22 dwarf irregular
galaxies within a distance of 7 Mpc.  Since the images came from a
multitude of different programs, the number of pointings and
filters varied from galaxy to galaxy. Table \ref{galaxypars}
lists, in order of integrated magnitude, the galaxies studied
along with the pointings and filters used for each. The footprints
of each of these pointings can be seen superposed on broad-band
ground-based images of each galaxy in Figure \ref{figfoot}.

The filters are part of the
{\it HST} flight system, and can be calibrated to the Johnsons/Cousin UBVRI
system using transformation formulas provided by Holtzman \et\ (1995).
F336W corresponds to a U-band filter, F439W to B, F555W to the V-band, and
F814W to an I-band filter.
When available, we combined multiple exposures
of the same field of view and same filter to remove cosmic rays, while
preserving photometric integrity.
Exposure times of individual images varied from 140 s to 2800 s.

All of the pointings in each galaxy were searched for clusters using criteria
of a minimum FWHM and magnitude.  The minimum
FWHM used was 2 pixels for the WF CCDs, which have a pixel size of
0.099\arcsec, and 2.5 pixels for the PC CCD,
which has a pixel size of 0.0456\arcsec.
These limits were just larger than the point spread function (PSF), insuring that
the object is resolved with respect to a single star.
This does mean that in the more
distant galaxies our lower bound was intrinsically much bigger than in near galaxies,
and this trend can be seen in the resulting
FWHM of the clusters found.
We also imposed a magnitude minimum.
This was chosen to be the V magnitude of a cluster of mass 10$^4$ M\solar\
and an age of 1 Gyr. This corresponds to roughly an old populous cluster
or the low end of the range of globular cluster masses. These criteria
result in an M$_V$ limit of $-6.2$.
For all of our images, this magnitude limit is much brighter than
the photometric limit of the images and so all objects
brighter than this limit would be found (but see discussion of other complications
below). The images were searched by doing radial profiles on all objects
to well below our brightness cut-off to make sure that we were getting
all objects that fit our criteria. The FWHM of the clusters were determined
from the radial profile.

We performed photometry on all the clusters that fit these criteria (except
for cases of blatant cosmic ray interference) using an aperture that
contained the visible region of the cluster, usually about twice the FWHM
but occasionally more.
To subtract the sky contribution from the cluster luminosity, we used the
mode of sky brightness values in an annulus measured around the cluster.
The results of the photometry were then calibrated to the WFPC2
system using the synthetic zero points of Holtzman \et\ (1995).

In order to deredden the photometry, we used STSDAS
simulations of blackbody spectra
through
the various filters to find the difference between expected and observed
magnitude values, since the reddening correction is a
function of the spectral energy distribution of the object.   The
extinction values used were an internal E(B$-$V) of 0.05 for each
galaxy plus individual values for foreground extinction due to dust within
the Milky Way depending on the location of galaxy (values taken from
Burstein \& Heiles 1984).
A single reddening was used for all clusters in a galaxy.

In the case where the
filter F336W was used, we further corrected for the red leak
in the filter by again using STSDAS simulations. The correction
was a function of an observed color such as F555W$-$F814W.
The final step was to convert the flight magnitudes to the Johnsons/Cousins
system as specified in the formulas of Holtzman \et\ (1995).
In the case of some galaxies where the only
available filters were F606W (V$+$R) and F814W (I), we compared the
observed F606W$-$F814W color to
blackbody spectra as seen through F606W, F814W, R and I
to get R$-$I from F606W$-$F814W.  We used the conversions from
Holtzman \et\ (1995) to get I and then
solved for the R magnitude.

Our objects should be considered {\it candidate} compact clusters.
Among other problems, we may have imposters in the ranks. Given
the low dust density of dwarf irregular galaxies, it is often
possible to see background galaxies in the images, and bright
elliptical galaxies can be hard to distinguish from red clusters.
At the outer edges of galaxies they can be identified but towards
the center, the more likely location for clusters, the subtle
distinction between the two poses a more serious risk. Rather than
discard potential candidates, we chose to include all and accept
that some may be background galaxies. However, we did exclude
resolved objects that were located in the outer parts or beyond
the optical galaxy as most likely background galaxies. In addition
we cannot be certain that all objects are clusters---a collection
of stars formed in a single star-forming event. Some may be a
chance superposition of several stars along the line of sight.
Because of our desire to include faint old populous clusters, our
magnitude limit overlaps with the brightest stars found in
galaxies. A few of these stars in close proximity could fall
within our constraints.

In addition to the possibility of imposters, we also have problems that would
lead us to miss clusters. Particularly,
clusters that are saturated on
the available CCD images and not already known to be a cluster
would not make it onto our list.
Also, for half of the
galaxies searched, there was only one exposure in each filter, leaving it
vulnerable to the mercies of cosmic rays. Cluster candidates that were
found in one filter occasionally could not be measured in another due to
intrusive cosmic ray interference.  Also, there were a few candidates
annoyingly located at the boundaries of the chip, making it impossible to
get a good sky value for that area and forcing us to discard the object.
In addition, the {\it HST} images only covered a fraction of the optical
galaxy in most cases. The fraction of the area covered
within D$_{25}$, the diameter
at which the galaxy drops to a B-band surface brightness of 25 mag arcsec$^{-2}$,
is listed in Table \ref{galaxypars}.

\section{Results}

Of the 22 galaxies searched, we found candidate compact clusters
in 8 of them.
Table \ref{tabcld} shows the cluster demographics
in these galaxies.
In Tables 4--11 we present parameters of each cluster found.
Figure \ref{figfwhm} shows a
histogram of the measured FWHMs for all of the clusters. Most
common is a FWHM of between 4 and 6 pc. Thus, these are indeed
compact clusters, and comparable to those seen in other galaxies.

In order to determine an age for each cluster, we compared the
integrated cluster photometry to
cluster evolutionary models of Leitherer \et\ (1999). Their
models chart the colors of a
cluster with a mass of 10$^6$ M\solar over time, assuming a
Salpeter (1955) stellar initial mass function (IMF) of 0.1 to 100
M\solar.  The models are
dependent on the metallicity $Z$ of the host galaxy.
We estimated $Z$ from
the more commonly available oxygen abundance found from nebular emission.
Most galaxies have metallicities near $1/5$ Z\solar.
Studies of clusters in NGC 1569 and NGC 4449 have found that the ensemble
of clusters were fit better with cluster evolutionary tracks of a somewhat
higher metallicity than one would have predicted from the nebular oxygen abundance
(Hunter \et\ 2000; Gelatt, Hunter, \& Gallagher 2001). For this reason,
we bracketed the data between
models of two $Z$ where possible to allow better comparison.
However, we have chosen to use the higher metallicity model to
estimate the age of the clusters.
The clusters from each galaxy are shown on
color-color or color-magnitude plots along with the models
in Figures \ref{fign4214} to \ref{figsexa}.

Age estimates from comparison with cluster evolutionary models are
uncertain in any event. This is especially true in small clusters
where stochastic effects due to the small numbers of stars
can wreak havoc. This problem is discussed and simulated by
Girardi \& Bica (1993; see also Santos \& Frogel 1997, Brocato \et\
1999). Their simulations show a scatter of several tenths in
UBV colors. In addition to this problem, for many galaxies
we have only one color to work with. The additional information
of the second color is important in constraining an age.
For clusters with only one color we have simply noted where
the cluster color intersects the cluster evolutionary tracks
($\pm$0.05 mag).
In the cases where clusters
were too red or blue to fall near the evolutionary tracks, we
assigned the age of the
place on the tracks that is closest to the clusters;
these cluster ages are denoted by a colon to
mark their uncertainty.  Some clusters fell too far from the
tracks to assign an age at all.
In cases where clusters had a large range of possible ages and were within
a region of \ha\ emission, we assumed that it must be young and used the lower
age as a value for that cluster. For a few clusters where the
age predicted by the colors
were in conflict with the presence of nebular emission, we assigned a
young age consistent with the nebular emission.
In these cases, as well as others, variable extinction across the galaxies
may contribute to moving a cluster away from the cluster evolutionary
tracks. In a U$-$V versus V$-$I color-color diagram a reddening line
moves a cluster diagonally to the lower right (see Figure \ref{fign4214}).

The clusters are of different ages, and since clusters fade with time,
to compare absolute magnitudes, and hence relative mass, we must
compare at a fiducial age.
We have chosen a standard of 10 Myr, and used the models to determine
how much to correct the magnitude of each cluster.
When the photometry of a cluster corresponded to a range of possible ages in the model,
we calculated the range of magnitudes at 10 Myr for
those ages. For those clusters that fell unreasonably far from the evolutionary
tracks, we have not estimated a magnitude at 10 Myr and they are not counted
in our census.
The M$_V$ at 10 Myr of the collection of clusters are shown in Figure \ref{figmag}.

The terms super-star cluster and populous cluster are not
quantitatively defined. Hodge (1961) first used the term populous
cluster to refer to the compact clusters in the Magellanic Clouds.
The use of super-star cluster arose later to emphasize their
extreme nature (van den Bergh 1971).
Here, we have chosen to define these cluster categories
using the M$_V$ of the cluster predicted for an age of 10 Myr. We
have chosen to define a super-star cluster as a cluster with a
magnitude at 10 Myr of at least $-$10.5.  This is the absolute
magnitude that R136 is predicted to have at that age. The
criterion for a populous cluster is taken from Larsen \& Richtler
(2000). They used a limit of M$_V=-8.5$ for red objects
(U$-$B$>-0.4$) and an M$_V$ of $-$9.5 for blue clusters
(U$-$B$<-0.4$). Since we are using M$_V$ related to 10 Myr, we use
an M$_V$ lower limit of $-$9.5 for these clusters. For comparison,
we collect properties from the literature of known super-star
clusters found in other dwarf irregular galaxies in Table
\ref{tabssc}.

We have determined the number of clusters N$_{-9.5}$ with M$_V$ at
10 Myr brighter than $-$9.5. This includes populous and super-star
clusters. This number divided by the area of the galaxy surveyed
(within D$_{25}$) is the number density ND$_{-9.5}$. Similarly,
the number of super-star clusters is N$_{-10.5}$ and the number
per unit area is ND$_{-10.5}$. These values are given for each
galaxy in Table \ref{tabcld}. We also include in this table
several dwarf irregular galaxies studied by others: IC 10 (Hunter
2001), NGC 4449 (Gelatt et al.\ 2001), NGC 1569 (Hunter \et\
2000), and VIIZw403 (Lynds \et\ 1998). IC 10 and NGC 1569 are
starburst galaxies
(Massey \& Johnson 1998; Gallagher, Hunter,
\& Tutukov 1984), NGC 4449 has been disturbed by an interaction
in the past 1 Gyr (Hunter \et\ 1998). Of these, only NGC 4449 and
NGC 1569 contain populous and super-star clusters.

Classification as a populous or super-star cluster depends critically on the
estimate of age since this in turn determines the prediction of M$_V$ at 10 Myr.
An old cluster has a large correction to M$_V$, and such an age estimate
misapplied could turn a very modest cluster into a super-star cluster.
Our uncertainties over ages in many cases are reflected in the range in
numbers given in Table \ref{tabcld}.
The lower number includes only clusters with a single age estimate, as given
in Tables 4--11. The larger number in the range includes clusters that would
fall into our magnitude bin {\it if} the older age is the correct one. In most
cases, the uncertainty in age reflects the multi-valued nature of the models
for a single color; a second color would improve the age estimates.
In the discussion and plots below, we conservatively use only the lower numbers
in the ranges in Table \ref{tabcld}.

\subsection{NGC 4214}

This starburst galaxy was a jackpot.
We have identified 29 cluster candidates.
In Figure \ref{fign4214} we can see that there are several clumps
of clusters. One clump (clusters 5, 12, 18, 28) is far too blue in
(V$-$I)$_0$ for the models. Contamination of V by \ha\ emission
is a possibility. We can see diffuse light around cluster 5
and cluster 18 is sitting in nebulosity. On the other hand,
we can see nothing unusual with clusters 12 and 28.
We have assumed that these clusters are young, but because their
colors are so unusual, we have not determined M$_V$ at 10 Myr
and so cannot compare them to other clusters.
Another clump of clusters falls just redward of the tail end of the
cluster evolutionary track that ends at 1 Gyr.
We have assumed an age of 1 Gyr for these clusters
(clusters 1, 2, 4, 13, 17; clusters 16 and 20 are assumed young because
they sit in \ha\ emission regions).
This region of the color-color diagram is also where one would expect
Milky Way globular clusters to sit (Reed 1985), so these clusters
could be older than 1 Gyr.
However, this location would
also be the place where we would expect to find background early-type galaxies
(Poulain 1988), so there is also a higher probability of contamination
from non-galactic sources.
Finally, there is a clump
of clusters that fall a little redward of the young leg of the evolutionary
track in (U$-$V)$_0$ but blueward of the older leg in
(V$-$I)$_0$. We have assigned ages according to which branch the cluster
falls closest, or in some cases our dual assignment reflects our uncertainty.

Leitherer \et\ (1996) had
identified a possible embedded super-star cluster of 4--5 Myr
in the \HII\ complex at the center of the galaxy.
However, we have identified another 6 which
qualify as super-star
clusters (clusters 1, 2, 4, 10, 13, 17)
{\it if} their age estimates are correct.
All are relatively old. If the older age is the correct one for clusters
3, 21, and 24, they also would fall in the super-star cluster camp.

Clusters 15, 16, 26, and 27 qualify as populous clusters, but only cluster 16
is young.
If the older age of cluster 11 is appropriate, it would also qualify
as a populous cluster.

This fruitful harvest may be due to the properties of the host galaxy; NGC
4214 is the brightest and largest galaxy surveyed and has a high
star formation rate.
Ma\'iz-Apell\'aniz (2000) classes NGC 4214 as a starburst system.
The clusters that we have found are
only a lower limit to the numbers since
the area surveyed by {\it HST} represents only about 20\% of the area
defined by its D$_{25}$ diameter. On the other hand, the 20\% of
the area searched
was in the center, near
or in the large \HII\ complexes, so the area searched is the most likely to contain
clusters.

\subsection{NGC 2366}

NGC 2366 is the second brightest galaxy surveyed, but sadly produced
nowhere near the spectacular results of its rival, NGC 4214.  Only about
38\% of this galaxy was surveyed, although again the pointing was in a likely part
of the galaxy. It does contain 3 clusters, but none merits the populous
cluster classification. One cluster, labeled B after Drissen \et\ (2000),
is located in the \HII\ complex numbered II by Drissen \et\ and
given an age of 3 Myr.

\subsection{DDO 50}

DDO 50 is the next brightest galaxy, and also has 3 clusters.
Cluster 3 in Table \ref{tabd50} is potentially a super-star
cluster depending on its age, which is not well constrained, but
if the young end of the range is appropriate, it would fall
below even our populous cluster cut-off.
The other two clusters did not fit the model
well enough to estimate their magnitude at 10 Myr.

The cluster census is less certain  here, since only 20\% of the D$_{25}$ area of
DDO 50 was surveyed by {\it HST} and the area surveyed was towards the
edge of the D$_{25}$ perimeter. In addition, the proximity of the
clusters to the edge of galaxy increases the risk that they could be
background galaxies instead.

\subsection{NGC 1705}

This galaxy, also a starburst, contains a known super-star cluster,
previously studied by Melnick, Moles \& Terlevich (1985)
and O'Connell, Gallagher \& Hunter (1994). This cluster,
labeled ``A,'' is
included in Table \ref{tabn1705}.
We used data from O'Connell \et\ (from WFPC1)
since the cluster saturated all frames used in
this survey. Fifteen other cluster candidates were found in this galaxy.
In Figure \ref{fign1705} we see that most of the clusters fall into
two clumps. One clump (clusters 10--12 and 14) falls around 7 Myr
on the cluster evolutionary track. The other clump falls redward
of the 1 Gyr end of the track. These clusters we have tentatively
assigned an age of 1 Gyr although contamination by background galaxies
is possible for these colors. Clusters 1 and 15 do not appear
in Figure \ref{fign1705} because they fall beyond the red corner
of the plot, far from the evolutionary track.

The majority of the clusters
were found surrounding the super-star cluster.
While in other
galaxies the clusters are relatively separated, here 10 of the clusters
were found encircling the super-star cluster.
These clusters
are identified in Figure \ref{fign1705around}.
The {\it HST} footprint is helpfully positioned with the PC chip centered on
the super-star cluster; since the PC chip has the highest resolution,
this proved useful in detecting the multitude of clusters around the
center one.
Eight of these clusters (clusters 7--14) are arranged in a semi-circle
within 13\arcsec$=$325 pc of the super-star cluster. Two others
(clusters 4 and 5) are somewhat more detached but still within
the same hemisphere. These two clusters plus two of the others were
assigned an age of 1 Gyr because they are very red. The others have
ages of 7 or 15 Myr. The age of the super-star cluster itself
is 7--20 Myr. If the age of 1 Gyr for the red clusters is
correct, their M$_V$ corrected to 10 Myr would place three
of them in the category of populous clusters (clusters 4, 5, and 9)
and one would be a super-star cluster (cluster 7). None of
the young clusters, which presumably formed at the same time
as super-star cluster A, are bright enough to fall into the
populous cluster category.

Of the rest of the clusters in the galaxy, only three red clusters
that we have dated at 1 Gyr are comparable to populous
(cluster 2) or super-star clusters (clusters 3 and 6).
The fact that only old clusters, with a large correction for fading
with time, have M$_V$ at 10 Myr bright enough to be classed as populous or
super-star clusters does make us suspicious that they
are background galaxies.

Nevertheless, it seems an impressive number of clusters to find
in a galaxy that is so much smaller than NGC 4214, the only other galaxy to
have clusters in comparable abundance.
The cluster density is the highest in our survey sample, exceeded only by
NGC 1569 (Hunter \et\ 2000).
However, this density is exaggerated due to the dense
packing around the super-star cluster, and probably does not apply to the
entire galaxy.

\subsection{DDO 168}

As seen in Table \ref{tabd168}, DDO 168 has 3 clusters in the area
surveyed, 46\% of the area defined by
D$_{25}$. Only about half of the {\it HST} footprint was within the
galaxy, and those parts of the CCDs that were exterior contained quite a
few identifiable background galaxies.  If we assume that these
background galaxies, our
own species of cosmic vermin, exist in the same density through the galaxy
as around it, the risk that the clusters found are actually galaxies in
disguise is a little higher and the reality of the clusters a little more
doubtful. One of the clusters, cluster 1, if the older age is appropriate
would be classed as a populous cluster.

%\subsection{IC 1613}
%
%Despite extensive searching, IC 1613 proved to contain only one candidate
%for cluster status (Table \ref{tabic1613}), and it unfortunately did not
%fit the model cluster
%evolutionary track well enough to allow us to estimate its magnitude at 10
%Myr. However, only about 5\% of the galaxy was surveyed by {\it HST}, so
%the cluster density is fairly high and there may be much more still
%hidden.

\subsection{DDO 165}

In the survey of this galaxy, again, there is a substantial sample of
background galaxies, making the 4 clusters found slightly doubtful
(Table \ref{tabd165}).  We surveyed 60\% of
the D$_{25}$ area and excluded the parts of the CCDs outside the
optical galaxy. The ages of these clusters are uncertain because
we only had one color and several ages would be consistent with the
colors. If the older age is correct, two of the clusters would
be super-star clusters and one, a populous cluster.

\subsection{WLM}

WLM has the advantage of already having a known globular cluster,
studied by Hodge \et\ (1999). It was noticed next to WLM by
Humason, Mayall, \& Sandage (1956) and later found to be
associated through a common radial velocity. Determined by Hodge
\et\ to be very similar to our own Galactic halo globulars, this
is cluster 1 in Table \ref{tabwlm}. It is measured here to have an
M$_V$ of $-$8.9, close to the value of $-$8.8 quoted by Hodge \et\
using data from Sandage \& Carlson (1985). This globular cluster
is unusual in so faint a galaxy, the only one to exist in a dwarf
irregular galaxy with a M$_V$ fainter than $-16$ in the Local
Group, according to Hodge \et\ Minniti \& Zijlstra (1977) searched
for other red globular clusters in WLM and found none. The only
other compact cluster that we found in the 23\% of the D$_{25}$
area searched does not fit the cluster evolutionary models well
enough to calculate a magnitude at 10 Myr for comparison. However,
it is certainly smaller and fainter than the globular cluster.

\subsection{Sextans A [DDO 75]}

Sextans A has the dubious distinction of being the smallest and least
luminous galaxy in this survey to contain any identifiable compact clusters.
It has two compact cluster candidates, listed in Table \ref{tabsexa}.
Cluster 1 could be a small super-star cluster if the older age is correct.

\section{Discussion}

\subsection{General Trends}

Galaxies with high star formation rates and high luminosities
should have a high abundance of clusters. The top panels of Figure
\ref{figsfr} show the number density of populous and super-star
clusters, ND$_{-9.5}$, versus the galactic M$_B$ and the
integrated star formation rate normalized to the area of the galaxy
SFR$_{25}$.
The SFR$_{25}$ is the star formation rate determined from the \ha\ emission (Hunter
\& Elmegreen 2002)
and normalized to the D$_{25}$ area using the formula of Hunter \& Gallagher
(1986).
The open circles are the new galaxies studied here,
from Table 1 and the top of Table 2; the triangles are the
smallest galaxies in the study by Larsen \& Richtler (2000; i.e.,
NGC 1569 and NGC 1705, which are also in our sample); the plus
symbols are three of the dwarf irregulars in the literature from
the bottom part of Table 2 (NGC 4449, IC 10, and VIIZw 403); and
the cross is the WLM galaxy. For galaxies with a range in numbers
reflecting our uncertainties in age estimates, we have
conservatively chosen the lower end of the range, thereby
excluding clusters that would fall into our bin only if an older
age were correct.

There is a vague correlation between ND$_{-9.5}$ and the galaxy
brightness or integrated star formation activity. Most of the galaxies that
contain populous clusters are brighter than an M$_B$ of $-16$ and
have a $\log$ SFR$_{25}$ greater than $-2.5$. The galaxy WLM
differs because it has a very low current star formation rate
($\log$ SFR$_{25}$ of $-3.8$) and an M$_B$ of only $-14$, yet it
contains am old globular cluster. Presumably the star formation
rate in WLM was higher when the globular cluster formed. The
cutoffs suggested by Figure \ref{figsfr} are not the only
conditions for populous cluster formation because there are
galaxies with values above both cutoffs that have not formed any.
In particular, the starburst galaxy IC 10
(Massey \& Johnson 1998) listed in Table
\ref{tabcld} and the Blue Compact Dwarfs (BCD) VIIZw403 (Lynds
\et\ 1998) and IZw18 (Hunter \& Thronson 1995) do not contain
populous clusters.
IC 10 and VIIZw403 have log SFR$_{25}$ of $-1.5$
and $-2.2$, respectively (Hunter \& Elmegreen 2002), both above the lower limit
for cluster formation, but their
mode of star formation has been that of
a scaled-up OB association (Hunter \& Thronson 1995, Lynds \et\ 1998,
Hunter 2001).

\subsection{The Larsen \& Richtler study of populous clusters}

Larsen \& Richtler (2000) examined the properties of what
they call ``young massive clusters'' in 21 nearby spiral and
irregular galaxies supplemented with data from the literature for
another 10 objects.
Their detection criteria (M$_V=-8.5$ for blue objects
and M$_V=-9.5$ for red objects) include populous clusters as well
as super-star clusters, and, since the super-star clusters are
rare, their sample is dominated by populous clusters. They find
that the formation of populous clusters is a normal and on-going
process in galaxies with adequate star formation activity.

Larsen \& Richtler (2000) developed a parameter $T_L$(U) to
represent the current cluster formation activity in a galaxy;
$T_L$(U) is 100 times the ratio of the combined U-band
luminosities of the clusters to the U-band luminosity of the whole
galaxy. For late-type spiral galaxies, $T_L(U)$ correlates with
$M_B$ and SFR. The bottom panels of Figure \ref{figsfr} show
these correlations. The small dots are the regular or starburst
spiral galaxies from Larsen \& Richtler. The smallest two galaxies
in their survey (NGC 1705 and NGC 1569) are plotted again as
triangles based on data from Tables 1 and 2; they are
considered separately because they have about the same diameters
as the dwarf galaxies considered here. The largest galaxies (NGC
7252, NGC 3921, NGC 3256, and NGC 1275) are plotted as squares.
The open circles and plus symbols are from our dwarf galaxy data,
as in the top diagrams. To keep all of the data points on the
plot, we let $\log T_L$(U)$=-2$ if a galaxy has no populous
cluster; most of our dwarfs are in this category.
The T$_L$(U) for our galaxies were determined from U-band photometry
for the galaxies from de Vaucouleurs \et\ (1991) and for the
clusters from the data presented here. In some cases a U filter
was not available for a cluster. In that case we used the available
colors and magnitude plus cluster evolutionary models (Leitherer \et\ 1999)
to predict M$_U$.
We do not account for the fact that the {\it HST}
fields do not cover the entire galaxy.

Larsen \& Richtler (2000) noted the trend of $T_L$(U) with M$_B$
and recognized that NGC 1705 and NGC 1569, the two triangles in
our plot, have magnitudes that are too faint for their $T_L$(U),
as shown also in our figure. Larsen \& Richtler's relationship
between $T_L$(U) and star formation rate is evident in the lower
right panel of Figure \ref{figsfr}. A bi-variate least squares fit
to the dots and squares (i.e., all of their galaxies except the
two smallest) has a slope of 0.97, as shown by the dashed line.

The dwarf irregulars in our survey do not always follow the SFR
correlation. Some of the dwarfs with high star formation rates
(NGC 1569, NGC 1705, NGC 4449) have populous clusters and follow
the trend, but most of the dwarfs do not. WLM currently fits the
trend, having $T_L$(U) near zero, but when it was younger the
globular cluster would have been much brighter, possibly
dominating the U-band emission from the whole galaxy
unless the SFR over the entire galaxy was
extraordinarily high at the same time.

\subsection{Super-star Clusters}

Super-star clusters, with M$_V<-10.5$ at a fiducial age of 10 Myr, are rare in
normal spiral galaxies, yet five nearby dwarf irregular galaxies
were previously found to contain them (NGC 1569, NGC 1705, NGC
4449, LMC, and WLM), and one additional host was found here: NGC
4214 (Table \ref{tabssc}). Even more galaxies in Table
\ref{tabcld} could have super-star clusters if an older, but
highly uncertain, age is correct (we will not count these here).
There are also 4 dwarf galaxies, listed in Table \ref{tabssc} with
a "?" mark, that are believed to contain an {\it embedded}
super-star cluster.
The properties
of these clusters,
including their compactness,
are inferred from reradiation from the surrounding nebula rather than
from direct optical observations of the clusters, so their classification
as super-star cluster, rather than a very large OB association,
is less certain than if the cluster were visible.
The table suggests
that galaxies which host a young super-star cluster are all
brighter than M$_B=-16$.  Most of the super-star clusters are
young, with the exception of the globular cluster in WLM. They are
also compact, with sizes and masses comparable to globular
clusters.

Table \ref{tabssc} lists any crucial status of the galaxy that
hosts a super-star cluster. Most of the galaxies show some sign of
being affected by an interaction with another galaxy, either
currently or in the past. Furthermore, most of the super-star
clusters are located near the centers of the galaxies (see also
Meurer et al.\ 1995). A few that are not located near the centers
are located at the end of a stellar bar. The centers of
interacting galaxies are special places since interactions can
drive gas into this region (Noguchi 1988). Similarly, streaming
motions around bars can pile the gas up there (Elmegreen \&
Elmegreen 1980). This correlation between the presence of
super-star clusters and large-scale galactic disturbances suggests
that dwarf irregulars need a special mechanism, like an external
perturbation, internal stellar bar, or shell-shell interaction
(Chernin, Efremov, \& Voinovich 1995; Fukui 1999), to concentrate
the gas enough to form a massive compact cluster.

The exception to all of this is WLM. With an integrated M$_B$ of
only $-14$, it nevertheless contains a globular cluster. When the
globular cluster was 10 Myr old, it would have had an M$_V$ of
order $-14$, rivaling the brightness of the entire galaxy as seen
today. Furthermore, WLM does not show any particular sign of
having been disrupted by an interaction 15 Gyr ago or of having
gone through an extraordinary starburst then, although our ability
to probe the star formation histories that far back is limited.
The existence of this globular cluster in such a small galaxy and
its rarity does suggest that something extraordinary happened to
WLM nearly 15 Gyr ago when the galaxy was first forming.

Other dwarf irregular galaxies are undergoing starbursts that do
not show evidence of compact star formation and super-star
clusters.  IC 10, for example, is a starburst galaxy (Massey \& Johnson 1998)
that has not
produced anything but small clusters and OB associations
at least in the portion imaged with {\it HST} (Hunter
2001). VIIZw403 and IZw18 are BCDs that have only made large OB
associations (Hunter \& Thronson 1995, Lynds \et\ 1998). Neither
the conditions that produce a starburst nor the starburst itself
are sufficient to produce clusters of such an extreme nature.

We do find a rather gregarious nature to super-star clusters. The
super-star cluster in NGC 1705 has a collection of 10 smaller
compact clusters arranged in a semi-circle around it. Six of these
have an age that is similar to that of the super-star cluster.
There are also 10 compact clusters surrounding the youngest of the
two super-star clusters in NGC 1569 (Hunter \et\ 2000). Again, at
least 5 of these have an age comparable to that of the super-star
cluster and another 3 have ages only $\sim$10 Myr older. There are
also OB associations formed over the past 10 Myr arrayed around
the super-star cluster R136 in the LMC that is itself 2 Myr old.
It appears that these super-star clusters have formed from very
large gas complexes that produced compact clusters and OB
associations as associated debris. Star formation in such
complexes appears to extend over a time-scale of roughly 10 Myr.
However, not all such complexes form super-star clusters
(Kennicutt \& Chu 1988). Constellation III in the LMC, for
example, is an older complex that formed stars over a scale
comparable to the 30 Doradus nebula in which R136 sits. However,
only a few populous clusters have formed in Constellation III
(Dolphin \& Hunter 1998).

\section{On the Origin of the Correlations}
\subsection{Local and Global Effects}
\label{sect:origin}

The gregarious nature of super-star clusters is consistent with
the previous observation that stars form primarily in large
regions where the gas column density exceeds a threshold value for
strong gravity and a cool phase of HI (Skillman 1987; van Zee et
al. 1997; Hunter, Elmegreen, \& Baker 1998).  In dwarf irregular
galaxies, these pockets of high pressure are very localized
because most of the interstellar medium is at low pressure (Elmegreen \& Hunter
2000).  As a result, dense clusters form with other clusters and
OB associations in islands of high pressure and high column
density, with little activity between. Spiral galaxies are
different because they have a pressure that exceeds the threshold
throughout the main disk, so star formation is distributed as
well.

The similarities and differences between the formation of dense
massive clusters in dwarf and spiral galaxies gives us some
insight into the correlation between $T_L(U)$ and SFR$_{25}$. There is
generally a relation between star formation rate per unit area,
${\rm SFR_{25}}$, and gas column density, $\Sigma$, that spans a wide
range of conditions, i.e., ${\rm SFR_{25}}\propto\Sigma^{1.4}$
(Kennicutt 1998). There is also an expectation that column density
correlates with pressure too, from the self-gravity of the gas
layer, giving $P\sim\left(\pi/2\right)G\Sigma^2$.  Thus $P\propto
{\rm SFR_{25}}^{1.4}$.  The formation of a dense cluster also depends
on pressure. All self-gravitating objects have a pressure that
scales with the square of the mass column density, so both dense
clusters and molecular cloud cores will have high pressures.  In
the case of a cluster, this pressure is kinematic, from the
motions of stars, but it reflects the high gas pressure in the
cloud core where it formed.

Two equations may be used to relate cluster mass to pressure,
given the characteristically high density that is used to define a
cluster.  These are the virial theorem, $c^2\sim0.2GM/R$, 
where $c$ is the velocity dispersion and $R$ is the radius of
the cluster,
and the
density-mass-radius relation, which may be converted into a
pressure-mass-radius relation for average internal pressure
$P_{int}\sim0.1GM^2/R^4$.  These two equations give a relation
between mass, pressure and density:
\begin{equation}M\sim6\times10^3\;M_\odot\left(P_{int}/10^8\;{\rm
K\;cm^{-3}}\right)^{3/2} \left(n/10^5\;{\rm
cm}^{-3}\right)^{-2},\label{eq:mass}\end{equation}
where $n$ is the gas volume density.
The normalization for this
relation comes from the properties of the molecular core near the
Trapezium cluster in Orion (Lada, Evans \& Falgarone 1997). The
cluster mass derived in this way is a maximum for two reasons:
smaller clusters can fragment out, and the cluster mass function
may not sample out to this high value if there are not many
clusters overall.  Also, larger clusters can form by coalescence,
but their pressures will not reflect the initial cloud core
conditions. From this relation we get $M_{max}\propto
P_{int}^{3/2}$ for a given stellar density in a cloud core.
Virialized regions with lower average densities can form
higher-mass star-forming regions for a given pressure, but these
regions will be called OB associations and not clusters because of
their lower densities. If the internal pressure $P_{int}$ scales
with the ambient pressure, $P$, then the maximum cluster mass
$M_{max}\propto {\rm SFR_{25}}^2$.

Now we sum all the clusters to find the the total cluster mass,
which is proportional to the total cluster L(U). 
If $n(M)dM=n_0M^{-2}dM$ is the cluster
mass function (e.g., Zhang \& Fall 1999), then the integral over
$n(M)$ from $M_{max}$ to infinity gives the normalization factor
$n_0$, i.e. $\int_{M_{max}}^{\infty}n_0M^{-2}dM=1$, or
$n_0=M_{max}$. With this normalization, the total cluster mass is
$M_{tot}\propto M_{max}\ln\left(M_{max}/M_{min}\right)$, which
depends only weakly on the minimum cluster mass, $M_{min}$.
Similarly, the total number of clusters, $N$, equals approximately
$M_{max}/M_{min}$, as may be determined from
$\int_{M_{min}}^{\infty}n_0M^{-2}dM=N$. For fixed $M_{min}$
defined by the survey, i.e. by the definition of a certain type of
star cluster, such as galactic, populous, super, etc., $M_{max}$
scales directly with the total number of clusters, $M_{max}\propto
N$, as found observationally by Whitmore (2000). The total mass in
clusters scales mostly with $M_{max}$ because the logarithmic term
is slowly varying. Thus $M_{tot}\propto M_{max}\propto {\rm
SFR_{25}}^2$.

The top left part of Figure \ref{fig:masslum}  shows the total
U-band luminosity of clusters brighter than M$_V=-9.5$  versus the
star formation rate per unit area, plotting normal or starburst
galaxies in Larsen \& Richtler (2000) as dots, 
the starburst dwarfs
NGC 1569 and NGC
1705 as triangles, and the four largest galaxies in their study
(NGC 7252, NGC 3921, NGC 3256, and NGC 1275) as squares.  The
total cluster luminosity is derived from their value of $T_L(U)$
and the U-band magnitude of the galaxy. The data for NGC 1569 and
NGC 1705 are taken from our tables here.  Other galaxies from our
tables are plotted as a circle (NGC 4214), a plus (NGC 4449), and
a cross (WLM) to be consistent with the symbols in Figure
\ref{figsfr}. The dashed line is a bi-variate least squares fit to
the dots and squares only, i.e., to the normal and large galaxies
in Larsen \& Richtler's study. The slope is $\sim1.9$, close to
the theoretical value of 2 given above. The dwarfs (circle, cross,
plus and triangles) have a similar trend but are displaced
downward in this plot because 
there are
fewer clusters and a smaller total cluster luminosity.

The bottom left part of Figure \ref{fig:masslum} shows the total
galaxy luminosity in U-band versus the star formation rate. The
dashed line is a fit to the dots and squares and has a slope of
$\sim1.2$.  This plot has an increasing trend because the galaxies
in Larsen \& Richtler's (2000) study all have about the same size
(see below). Thus the total U-band luminosity, which is related to
the star formation rate per unit area times the area, ends up
nearly proportional to the star formation rate per unit area.

The right hand side of Figure \ref{fig:masslum} shows the same
quantities normalized to the galaxy areas.  The dashed lines are
fits: the slope is 1.4 on the top and 0.5 on the bottom. The
correlations are much better for the normalized data than for the
un-normalized data. Dwarf galaxies that plotted below the mean
relations on the left in this diagram are brought in to a tight
correlation on the right.

The low value of the slope for galaxy $L(U)/{\rm Area}$ versus
$SFR_{25}$ is odd: it indicates that the average U-band intensity
of a galaxy does not scale with the star formation rate per unit
area (if it did, the slope would be 1).  There could be several
reasons for this. The U-band suffers more extinction than
H$\alpha$, which was used for $SFR_{25}$, and this extinction
increases systematically with column density and therefore with
$SFR_{25}$, making the $L(U)$ that escapes the galaxy increase
slower with the star formation rate than the intrinsic $L(U)$
(Hopkins et al. 2001).  Another possibility is that galaxies with
high star formation rates form stars for a longer time relative to
the average age of the HII region than galaxies with low star
formation rates. This seems counter-intuitive since high rates
should exhaust the gas supply faster than low rates, given the
$SFR_{25}\propto\Sigma^{1.4}$ relation for gas column density
$\Sigma$ (Kennicutt 1998). Nevertheless, the duration affects the
ratio between U-band light, which comes from stars younger than
$\sim100$ Myr (many of which are in the field -- Hoopes, Walterbos,
\& Bothun 2001) and H$\alpha$, which comes from stars younger than
$\sim10$ My inside HII regions. A third explanation is that the
average surface brightness of galaxies is about constant in B-band
(Freeman 1970), nearly independent of the star formation rate, so
the average surface brightness in U-band is somewhat constant too,
varying mostly because of a change in color with SFR$_{25}$. This
goes in the right direction for the correlation in Figure
\ref{fig:masslum}, but the actual color variations are not large
enough to explain the total effect.  Perhaps some combination of these
effects and others are responsible.  Whatever the reason, the
change in slope going from un-normalized to normalized $L(U)$ also
occurs for the total cluster $L(U)$, so the ratio of these two
quantities, which gives the parameter $T_L$, preserves its
near-linear correlation with $SFR_{25}$ regardless of
normalization.

The sizes of all the galaxies considered in this paper are shown
in Figure \ref{fig:sizes}.  The symbol types are the same as in
the previous diagrams.  The smallest galaxies in the Larsen \&
Richtler (2000) survey, NGC 1569 and NGC 1705 (triangles), are
about the same size as the dwarfs studied here (circles, plus, and
x symbols).  The dots have a narrow range of sizes, and the
squares are bigger, as defined. This separation between triangles,
dots, and squares in the Larsen \& Richtler data explains the
similar separation in the lower left part of Figure \ref{figsfr}. There
is a slight correlation between the star formation rate per unit
area and the galaxy size in the Larsen \& Richtler sample, but not
among the dots alone and the squares alone in this figure.  The
overall correlation gives approximately a galaxy
Area $\propto$ SFR$_{25}^{0.5}$, which explains the decreases in
slope of the total cluster 
$L(U)$ and the galaxy $L(U)$ versus
$SFR_{25}$ in going from un-normalized to normalized quantities in
Figure \ref{fig:masslum}.

The ratio of the quantities plotted at the top and bottom of
Figure \ref{fig:masslum}, multiplied by 100, is the parameter
$T_L(U)$.  The linear increase in $T_L(U)$ with SFR$_{25}$ found
by Larsen \& Richtler (2000) is the result of a squared dependence
of the numerator on SFR$_{25}$, as determined in part by {\it
local} properties of cluster formation in environments with
varying surface densities, pressures, and maximum cluster masses
(discussed above where $M_{tot} \propto {\rm SFR_{25}}^2$),
and a linear dependence of the denominator on SFR$_{25}$, as
determined by the narrow range of galaxy sizes in the sample.

In a dwarf irregular galaxy, the pressure may be large enough to
make a massive dense cluster in a localized region of intense star
formation, but because that region is small, the total sample of
all clusters in the galaxy may be too small to produce one of the
uncommon clusters at the high mass end of the mass spectrum. In
this case, the star formation rate per unit area can be high over
a small region, but the galaxy can be without a super-star
cluster. Most of the dwarfs studied here are in this category.
They account for the circles in lower right part
of Figure \ref{figsfr}.

In a large galaxy with a high star formation rate per unit area,
the maximum cluster mass is large and the number of clusters is
also large enough to sample out to around this maximum mass. Then
the parameter $T_L(U)$ represents the fraction of stars that is
still in bound clusters after the evolution time for UV light.
This fraction should depend on only the local conditions for star
formation and consequently defines a {\it maximum value for
$T_L(U)$ in galaxies of any size}. These local conditions suggest
that most stars reside in dense clusters for the first
$\sim10^5-10^6$ years (Carpenter 2000), and then disperse when
their clusters emerge from the cloud cores (Lada, Margulis, \&
Dearborn 1984). At an age of $\sim10$ Myr, the mass fraction of
stars that are still in bound clusters is only $\sim10$\%.  This
is the fraction of young stars in the form of bound clusters in
the Solar neighborhood (Elmegreen \& Clemens 1985). This fraction
slowly decreases with time for a given population as the clusters
continue to evaporate and disperse. In the $\sim100$ Myr time
during which most of the UV light from star formation comes out,
only the smallest clusters evaporate.

The super-star clusters studied here and by Larsen \& Richtler
(2000), which are more massive than $\sim2\times10^4$ M$_\odot$,
have long evaporation times and are hardly effected by disruption
in the first 100 Myr after they leave their cloud cores (this
situation changes in the nuclear regions, where even massive
clusters disperse rapidly -- Kim  et al. 2000). Thus the fraction
of star formation in the form of super-star clusters is $\sim10$\%
for surveys of high-P galaxies involving UV light -- as long as
the galaxy has a large enough mass to sample the cluster mass
spectrum out to the high value of $M_{\max}$ given by the
pressure. 
This is what is seen in the bottom right panel of Figure \ref{figsfr}
where T$_L$(U)$\sim10$.
The cluster fraction should be lower than 10\% among
low-mass clusters, i.e., in the mass range from $10^2$ to $10^3$
M$_\odot$, even for high-P galaxies, because these clusters
evaporate more quickly than high-mass clusters in the U-band
evolution time.  The cluster fraction should be lower in redder
passbands too because the longer evolution time gives more
clusters an opportunity to disrupt.

The fraction 
of stars in massive bound clusters 
is also lower in lower pressure galaxies, which means
galaxies with lower normalized star formation rates, because then
$M_{max}$ is small as a result of the small total number of
clusters and the small pressure. This decreasing fraction is what
is shown by the Larsen-Richtler relation 
in Figure \ref{figsfr} (lower right panel)
among galaxies that
have about the same area, to within a factor of $\sim3$.  When the
area is much smaller, as in the dwarf Irregulars, it becomes
physically difficult to form a massive cluster if $P$ is low, and
it becomes unlikely to form one even if $P$ is high.  Occasionally
a dwarf will do this if the random fluctuations that make dense
clouds happen to produce a rare massive one.

\subsection{The Size of Sample Effect}

The smallness of dwarfs makes the appearance of even an occasional
super-star cluster statistically unlikely. In
Figure \ref{fig:sos} 
we plot $L_V$ of the brightest star cluster in each galaxy. 
The bottom three panels
show the size-of-sample effect for the
galaxies in Larsen \& Richtler's (2000) survey, plotted as filled
symbols as before, and it shows deviations from this statistical
effect for the dwarf irregulars.   For the Larsen \& Richtler
sample, the galaxies with larger numbers of bright clusters,
$N_{-9.5}$, larger total cluster $L(U)$, and larger total star
formation rates ($SFR_{25}*{\rm Area}$) also have larger most-massive
clusters. This is the size-of-sample correlation found by Whitmore
(2000). The dashed lines in these three panels have a slope of
unity and are a reasonable approximation to the data.  This linear
dependence was explained in Sect. \ref{sect:origin} as a
size-of-sample effect for a cluster mass function that scales as
$n(M)dM\propto M^{-2}dM$.

To make Figure \ref{fig:sos}, we used most of the data in Larsen
\& Richtler along with the references they gave, but we used
Calzetti et al. (1997) for NGC 5253, ignoring the embedded sources
found by Turner et al. (1998). The two dwarfs in the
Larsen-Richtler sample (NGC 1569 and NGC 1705) are plotted as
triangles in Figure \ref{fig:sos}, with V-band luminosity, $L_V$,
normalized to the fiducial cluster age of 10 Myr. The dwarfs in the
present survey are plotted as circles (NGC 4214, NGC 2366,
He2-10), a plus (NGC 4449), and a cross (WLM), with normalized
$L_V$ too.  We used Drissen et al (2000) for NGC 2366, and Johnson
et al. (2000) for He2-10.  Luminosities are in $L_\odot$.

The dwarf Irregular galaxies in Figure \ref{fig:sos} do not follow
the linear relation of the larger Larsen \& Richtler (2000)
galaxies. {\it The largest clusters in the dwarfs are too big for
the number of other clusters present} (i.e., they lie to the lower
right in the bottom three panels). The normalization 
of the cluster luminosity 
to a common
cluster
age for the dwarfs does not affect this result, because often the
clusters in dwarfs are younger than 10 Myr and the normalization
to 10 Myr would make
them fainter than they are, and because many of the
brightest clusters in the Larsen \& Richtler sample have ages of
about this value anyway.

The implication of the anomalous positions for dwarf galaxies in
the bottom three panels of Figure \ref{fig:sos} is that the
cluster mass functions in these galaxies are not continuous power
laws. There is probably a gap between the masses of normal
clusters and the masses of the super-star cluster. This figure
illustrates the main point of this paper,
{\it that super-star clusters
may require special or fortunate circumstances to form in dwarf
galaxies, and when they do, they have anomalously large masses}.

The top panel in Figure \ref{fig:sos} shows the squared dependence
between  SFR$_{25}$ and the luminosity of the brightest cluster,
as predicted following equation \ref{eq:mass}.  The dashed line
has a slope of $1/2$ (so the square of the ordinate goes with the
first power of the abscissa), and the dotted line, which is not a
good match to the data, has a slope of unity, as in the other
panels.

Figure \ref{fig:sos} illustrates three distinct properties of the
maximum cluster mass
in a galaxy.  First, because this mass scales about
linearly with the total size of the system for spiral galaxies
($M_{max} \propto$ total number of clusters, as shown above),
the largest mass
is a random sample from the cluster mass function. It is
affected by whatever makes the mass spectrum of clouds and
clusters, which is presumably related to turbulence (Elmegreen \&
Efremov 1997), and it does not depend much on the details of the
star formation process. However,
there is a second property
of the maximum cluster mass: it
scales with the square of the normalized star formation rate,
forcing the total cluster luminosity to scale
with SFR$_{25}^2$ too (Fig.
\ref{fig:masslum}).  Thus, there is a local influence in the
star formation process, like the pressure discussed in Section
\ref{sect:origin}. 
The difference between the global and local
aspects of cluster formation are not evident
from the spiral galaxy cluster sample because
most of these galaxies have the same size.

The differences between the global and local affects show up
better after comparing the spirals and dwarf Irregulars.  This
brings us to the third property of clusters evident from Figure
\ref{fig:sos}. For small galaxies with large SFR$_{25}$, the
size-of-sample effect should not give a superstar cluster because
the total sample 
of clusters 
is 
small. Indeed, most of the dwarf galaxies
surveyed here do not have one, even if their normalized star
formation rates are as high as for the spiral galaxies in Larsen
\& Richtler's sample (see the circles in the bottom right panel of
Fig. \ref{figsfr}). However, when the SFR$_{25}$ is high, there {\it
are} high pressure regions that {\it can} form massive clusters
sometimes.  This occurs when the stochastic process of cloud
formation in a turbulent fluid happens to make a massive dense
cloud. When such a cloud forms stars, the resulting cluster will
be more massive than expected from the extrapolation of the
cluster mass function elsewhere in the galaxy. Then the galaxy
will be plotted in the lower right parts of the bottom three
panels in Figure \ref{fig:sos}.

\subsection{Differences in Cluster Formation between Spiral and
Dwarf Irregular Galaxies}

The correlation between
massive cluster formation in dwarf Irregular galaxies and
galaxy-wide disturbances suggests that
turbulence may have been elevated before star
formation began. Interactions agitate a galaxy, increasing the
velocity dispersion and leading to large scale flows (Kumai, Basu,
\& Fujimoto 1993; Elmegreen, Kaufman, \& Thomasson 1993).  These
flows compress the gas, systematically at first and
then randomly as the turbulence cascades, and they ultimately make the
clouds in which clusters form.

Other cluster formation mechanisms are probably at work too. We
already mentioned end-of-bar flows and shell collisions as
mechanisms for creating anomalously high pressures. The 30 Doradus
region in the LMC could have been made by either. In any case, the
final condensation of gas into a massive dense core requires the
strong and persistent action of self-gravity. {\it The lack of
shear is perhaps the biggest factor differentiating dwarfs from
spirals}. Gravitational instabilities in both the ambient medium
and globally perturbed regions generally form giant clouds on the
scale of the local Jeans' length, which can be hundreds of parsecs
to several kiloparsecs in size, depending primarily on the ambient
density. When there is little shear, these clouds retain their
shape and condense over time as the turbulent energy dissipates
and the angular momentum moves outward along magnetic field lines.
The result can be a cloud core that is both massive enough and
dense enough beneath the weight of the overlying material to form
a super-star cluster.

Gravitational instabilities are very different in large galaxies,
which always have shear in their main disks and may also have
stellar spiral waves. If there are no stellar spirals, then
gaseous instabilities form clouds as in dwarf galaxies, but these
clouds quickly shear into flocculent spiral arms. The reason for
this is that the shear time is comparable to the instability time
when the gas surface density is close to the critical value and
the rotation curve is approximately flat. Smaller instabilities
and smaller clouds form inside the shearing arms by parallel
flows (Elmegreen 1991; Kim \& Ostriker 2002).
As a result, the initial Jeans-mass cloud
fails to collapse into a single dense object. In galaxies with stellar
spirals, ambient gravitational instabilities form clouds primarily
in the arms where the density is high; the shear is low there, so
these clouds can take a globular shape forming the familiar
beads-on-a-string pattern. However, spiral arm clouds reside in
this low-shear environment for too short a time to be able to
condense monotonically into massive dense cores
(Elmegreen 1994; Kim \& Ostriker 2002). They only have
time to build up modest pressures, and as a result, make
small clusters and unbound associations with a wide range of
masses.

A similar argument about shear applies to the formation of giant
shells around OB associations: dwarf galaxies can have shells (as
in the LMC, Kim et al. 1999; Ho II, Puche et al. 1992; IC 2574
Stewart \& Walter 2000) that are much larger compared to their
galaxy sizes than the shells in spiral galaxies.  This is because
the pressure from star formation or cloud impacts
in a low-shear galaxy can drive expansion
for a long time without any distortion of the overall shape.
Coriolis forces are still present, but without shear, they only
twist the whole structure around in a self-similar fashion as it
expands outward. This well-known difference in shell morphology
between dwarf galaxies and spiral galaxies is analogous to the
proposed difference in cloud morphology.

Super-star clusters also form in the nuclear regions of some
galaxies where the shear is low again. Nearby examples of
embedded nuclear clusters with thousands of O-type stars are in
NGC 253 (Watson et al. 1996; Keto et al. 1999) and NGC 5253
(Turner, Beck \& Ho 2000). An important difference is that the
ambient pressure and density can be very high in galactic nuclei,
so the clusters that form by gravitational instabilities can be
even more massive than those in dwarf irregulars.

\section{Summary}

After examining {\it HST} images of 22 nearby dwarf irregular
galaxies, we have found compact star clusters in 8 of them. Among
these 8 we found 3 galaxies (NGC 4214, NGC 1705, WLM) that host
populous clusters, which we define as clusters with M$_V$ at a fiducial age of 10
Myr brighter than $-$9.5. These same three galaxies host
super-star clusters, which we define as clusters with M$_V$ at an age of 10
Myr brighter than $-10.5$. Another 4 galaxies in our survey might
contain populous clusters and another 3 could join the list as
containing super-star clusters {\it if} the older age estimate in
a range of ages consistent with the cluster's color is correct.
However, this seems unlikely. We supplement our survey with our
previous census of compact clusters in IC 10, NGC 1569, NGC 4449,
and VIIZw403. IC 10, NGC 1569, and NGC 4449 contain compact
clusters candidates, but only NGC 1569 and NGC 4449 contain
populous clusters, and both also contain super-star clusters.

>From these host galaxies, if we exclude WLM, we find a magnitude
cutoff of M$_B=-16$ for the formation of populous and super-star
clusters; fainter than this and a galaxy does not seem to form
these extreme star clusters with a high enough probability to have
detected any in our sample. This is consistent with Hodge \et's
(1999) claim that no dwarf irregular galaxy, except WLM, in the
Local Group with M$_V$ fainter than $-16$ contains a globular
cluster. There is also a minimum normalized star formation rate of
0.003 M\solar\ yr$^{-1}$ kpc$^{-2}$ for cluster formation.  These
limits, however, are not sufficient to mandate the development of
clusters; there exist bright enough galaxies with high star
formation rates that are barren. The dwarf irregulars also
contribute to the trends seen for spiral galaxies by Larsen \&
Richtler (2000): The contribution of populous clusters to the
U-band light of a galaxy increases as the integrated galactic
M$_B$ and normalized star formation rate increases. However, a few
of the dwarf irregulars also contribute to the scatter and, again,
there are examples of galaxies with high M$_B$ or star formation
rate and zero contribution from populous clusters.

We can explain these results in terms of a general framework for
understanding the Larsen \& Richtler (2000) $T_L(U)$ correlations,
combined with the small sizes of dwarf galaxies and the
observation that most dwarfs with super-star clusters look
agitated. The correlations follow partly from the
inter-dependencies of pressure, gas column density, and star
formation rate in galaxies that are all about the same size and
also large enough to sample the cluster mass spectrum out to the
maximum cluster mass that is likely to form. Small galaxies cannot
generally sample out this far, and will usually fall short of
producing a massive compact cluster. Small galaxies with global
perturbations can make massive bound clusters, however.  They may
do this by the same star formation processes that operate in
spiral galaxies, in which case the dwarfs with super star
clusters are stochastic fluctuations in the dwarf sample.
Or, the dwarf may produce super star clusters by a
mechanism that is unique to their type, such as a gravitational
instability that is rendered more effective by the low rate of
shear. Similar processes may occur in the nuclear regions of large
galaxies.

\acknowledgments

We thank S.\ Larsen for useful comments.
OHB is grateful to the National Science Foundation for funding the
Research Experience for Undergraduates program at Northern Arizona
University under grant 9988007 to Northern Arizona University and
to Kathy Eastwood for directing it. Support to DAH for this
research came from the Lowell Research Fund and grant AST-9802193
from the National Science Foundation.  Support for BGE came from
NSF grant AST-9870112 and NASA grant HST-GO-08715.05A.

\clearpage

\begin{figure}
%\epsscale{0.8}
%\plotone{hunterda.fig1a.eps}
\caption{{\it HST} footprints of galaxies we surveyed shown on
ground-based images.
Images are displayed in the
V-band wherever possible, but some are from the Digitized Sky Survey (DSS) and
the IC 1613 image is a U-band image. The different pointings are numbered if
there was more than one of a particular galaxy.
\label{figfoot}}
\end{figure}

%\clearpage

%\begin{figure}
%\figurenum{1}
%\epsscale{0.8}
%\plotone{hunterda.fig1b.eps}
%\caption{continued}
%\end{figure}

%\clearpage

%\begin{figure}
%\figurenum{1}
%\epsscale{0.8}
%\plotone{hunterda.fig1c.eps}
%\caption{continued}
%\end{figure}

%\clearpage

%\begin{figure}
%\figurenum{1}
%\epsscale{0.8}
%\plotone{hunterda.fig1d.eps}
%\caption{continued}
%\end{figure}

%\clearpage

\begin{figure}
%\epsscale{1.00}
%\plotone{hunterda.fig2.eps}
\caption{Number count of
FWHM of the star cluster profiles for all candidate star clusters
found in the {\it HST} images.
The FWHM are not corrected for the WFPC2
PSF.
\label{figfwhm}}
\end{figure}

\begin{figure}
\caption{Number count of the M$_V$ at an age of 10 Myr
for all measured clusters.
For clusters with a range in ages and hence range in
predicted M$_V$ we have computed the number counts assuming
the lower ages in all cases and assuming the upper ages in
all cases.
The M$_V$ at 10 Myr are inferred from the
cluster evolutionary models of Leitherer et al.\ (1999).
\label{figmag}}
\end{figure}

\begin{figure}
%\plotone{hunterda.fig4.eps}
\caption{Star clusters found in NGC 4214 are shown in a UVI color-color diagram.
The solid curve is an evolutionary track for a cluster with
instantaneous star formation, a metallicity of 0.008, and
a Salpeter (1955) stellar initial mass function from 0.1 M\protect\solar\
to 100 M\protect\solar\ (Leitherer \et\ 1999).
Ages 1--9 Myrs in steps of 1 Myrs are marked with x's along these lines;
ages 10, 20, and 30 Myrs are marked with open circles.
The evolutionary tracks end at 1 Gyr.
The clusters are labeled as they are in the Table \protect\ref{tabn4214}.
The arrow in the lower left of the plot is a reddening line for an
E(B$-$V) of 0.2.
\label{fign4214}}
\end{figure}

\begin{figure}
%\plotone{hunterda.fig5.eps}
\caption{Star clusters found in NGC 2366 shown in a BV color-magnitude diagram.
The curves are evolutionary tracks for a cluster with
instantaneous star formation, a Salpeter (1955) stellar initial mass function
from 0.1 M\protect\solar\ to 100 M\protect\solar\ (Leitherer et al.\ 1999).
The solid curve is their model for a metallicity of $Z=0.004$
and the dashed curve is for $Z=0.001$.
Ages 1--9 Myrs in steps of 1 Myrs are marked with x's along these lines;
ages 10, 20, and 30 Myrs are marked with open circles.
The evolutionary tracks end at 1 Gyr.
The evolutionary tracks are for a cluster of mass 10$^6$ M\protect\solar;
for clusters of other masses the lines would slide up or
down in the diagrams. The clusters are labeled as they are in
Table \protect\ref{tabn2366}.
\label{fign2366}}
\end{figure}

\begin{figure}
%\plotone{hunterda.fig6.eps}
\caption{As for Figure \protect\ref{fign2366} except for the clusters
in DDO 50. Here the filters are R and I.
\label{figd50}}
\end{figure}

\begin{figure}
%\plotone{hunterda.fig7.eps}
\caption{As for Figure \protect\ref{fign4214} except for the clusters
found in NGC 1705. The colors here are (B$-$V)$_0$ and (V$-$I)$_0$.
\label{fign1705}}
\end{figure}

\begin{figure}
\caption{NGC 1705 super star cluster A and its surroundings are shown.
The image is the PC CCD of the F555W exposure. The compact clusters
surrounding cluster A are circled and identified according to the
numbers in Table \protect\ref{tabn1705}.
\label{fign1705around}}
\end{figure}

\begin{figure}
%\plotone{hunterda.fig9.eps}
\caption{As for Figure \protect\ref{fign2366} except for the clusters
found in DDO 168. Here the filters are R and I, and the
metallicities are $Z=0.004$ and 0.008.
\label{figd168}}
\end{figure}

\begin{figure}
%\plotone{hunterda.fig10.eps}
\caption{As for Figure \protect\ref{fign2366} except for the clusters
found in DDO 165. Here the filters are R and I, and the
metallicities are $Z=0.004$ and 0.008.
\label{figd165}}
\end{figure}

\begin{figure}
%\plotone{hunterda.fig11.eps}
\caption{As for Figure \protect\ref{fign2366} except for the clusters
found in WLM. Here the filters are V and I.
\label{figwlm}}
\end{figure}

\begin{figure}
%\plotone{hunterda.fig12.eps}
\caption{As for Figure \protect\ref{fign4214} except for the
clusters found in Sextans A. Here the filters are BVI.
\label{figsexa}}
\end{figure}

%\clearpage

\begin{figure}
%\plotone{hunterda.fig13.eps} 
\caption{Top panels: Number density
of the star clusters in each galaxy with M$_V$ at the fiducial age
of 10 Myr brighter than $-9.5$ plotted against the galactic M$_B$
and star formation rate per unit D$_{25}$ area (SFR$_{25}$). The
open circles are dwarfs from Table 1 and the top of Table 2; the
triangles are NGC 1569 and NGC 1705; the plus symbols are NGC
4449, IC 10, and VIIZw 403; and the cross is the WLM galaxy.
Bottom panels: Ratio of the luminosity of the clusters to the
galactic luminosity in the U-band, after Larsen \& Richtler
(2000). Galaxies with values of 0 for T$_L$(U) are plotted at a
value of $-2$ on the log plot. The small dots are from Larsen \&
Richtler; their largest galaxies (NGC 7252, NGC 3921, NGC 3256,
NGC 1275) are closed squares. The dashed line is a fit to the dots
and squares. \label{figsfr}}
\end{figure}

\begin{figure}
%\plotone{hunterda.fig14.eps} 
\caption{Left: The total U-band
luminosities of populous and super-star clusters (top) and the
galaxy U-band luminosities (bottom) are plotted versus the
normalized star formation rates for dwarf galaxies studied here
and other galaxies in the Larsen \& Richtler (2000) sample. The
symbol types are the same as in Figure 13. The dwarfs (circle,
plus, cross and triangles) lie below the spiral galaxies because
the dwarfs are smaller. Dashed lines are bi-variate least-square
fits to the normal and large galaxies in Larsen \& Richtler's
sample (dots and squares).  The fit at the top has a slope of 1.9,
and the fit at the bottom has a slope of 1.2. Right: Normalized
cluster and galaxy luminosities are plotted versus the normalized
star formation rates. The dashed lines are fits to the dots and
squares with slopes of 1.4 (top) and 0.5 (bottom).}
\label{fig:masslum}
\end{figure}

\begin{figure}
%\plotone{hunterda.fig15.eps} 
\caption{Galaxy areas defined by $\pi
R_{25}^2$ are plotted versus the normalized star formation rates
of all the galaxies included in this paper.  The symbols represent
galaxy samples, as in the other figures.} \label{fig:sizes}
\end{figure}

\begin{figure}
%\plotone{hunterda.fig16.eps} 
\caption{The size-of-sample effect is
shown in the bottom three panels by plotting the number of
clusters brighter than $M_V=-9.5$, the total cluster U-band
luminosity, and the total star formation rate, versus the V-band
luminosity of the brightest cluster extrapolated to 10 My cluster
age. Luminosities are measured in L$_{\odot}$. The total star
formation rate is in M$_\odot$ yr$^{-1}$. Larger samples have
larger brightest clusters for normal spiral galaxies, making the
linear relation shown by the dashed line and derived in the text.
Dwarf galaxies with super-star clusters (circles, triangles, + and
x symbols) do not fit this trend because their most massive
compact clusters are brighter than what is expected from their
small sizes.   The star formation rate per unit area, in M$_\odot$
yr$^{-1}$ kpc$^{-2}$, is plotted versus $L_V$ in the top panel.
The dashed line in the top panel has a slope of 1/2 and is a
reasonable approximation to the data, making the brightest cluster
luminosity proportional to the square of the normalized star
formation rate. The dotted line has a slope of unity. }
\label{fig:sos}
\end{figure}

\clearpage

\begin{deluxetable}{lclcccccc}
\tabletypesize{\scriptsize}
\tablewidth{0pt}
\tablecaption{{\it HST} Images and Galaxy Parameters \label{galaxypars}}
\tablehead{
\colhead{} & \colhead{Number of} &
\colhead{} & \colhead{} &
\colhead{D$_{\rm 25}$} &
\colhead{$\log$ SFR$_{\rm 25}$} & \colhead{D} &
\colhead{Area Searched} \\
\colhead{Galaxy} & \colhead{Pointings} &
\colhead{Filters} &
\colhead{M$_B$} & \colhead{(kpc)} &
\colhead{(M\protect\solar/yr/kpc$^2$)} & \colhead{(Mpc)} &
\colhead{(Percent)}
}
\startdata
NGC 4214 & 4 & F336W,F555W,F814W & $-$18.37 & 11.9 & $-$2.42 & 4.8 & 19\\
NGC 2366 & 2 & F439W,F547W,F606W,F814W & $-$16.51 & 8.1  & $-$2.71 & 3.4 & 38\\
DDO 50   & 1 & F606W,F814W & $-$16.25 & 7.4  & $-$2.90 & 3.2 & 20\\
NGC 3109 & 1 & F606W,F814W & $-$16.13 & 9.5  & $-$3.16 & 1.7 & 36\\
NGC 1705 & 1 & F439W,F555W,F814W & $-$16.12 & 2.8  & $-$1.70 & 5 & 59  \\
NGC 6822 &15 & F336W,F439W,F555W,F814W & $-$15.28 & 2.3  & -2.13 & 0.5 & 40 \\
DDO 168  & 1 & F606W,F814W & $-$15.23 & 3.7  & $-$3.07 & 3.5 & 46\\
DDO 63   & 1 & F606W,F814W & $-$15.16 & 4.0  & $-$3.28 & 3.8 & 42\\
IC 1613  & 2 & F439W,F555W,F814W & $-$14.57 & 3.3  & $-$3.26 & 0.7 & 5\\
DDO 165  & 1 & F606W,F814W & $-$14.52 & 2.6  & $-$3.75 & 2.6 & 60\\
WLM      & 2 & F555W,F814W & $-$14.25 & 3.4  & $-$3.79 & 1.0 & 23\\
Sextans A& 2 & F439W,F555W,F814W & $-$14.18 & 2.1  & $-$2.64 & 1.5 & 46\\
Mrk 178  & 1 & F606W,F606W & $-$14.02 & 1.7  & $-$2.11 & 4.6 & 100\\
DDO 167  & 1 & F606W,F814W & $-$12.63 & 1.2  & $-$3.07 & 3.7 & 100\\
DDO 53   & 1 & F606W,F814W & $-$12.11 & 0.8  & $-$2.51 & 1.8 & 85\\
DDO 216  & 1 & F439W,F555W,F814W & $-$12.09 & 1.5  & $-$4.45 & 1.0 & 52\\
DDO 69   & 1 & F439W,F555W,F814W & $-$11.80 & 1.2  & $-$4.03 & 0.8 & 53\\
DDO 155  & 1 & F439W,F555W,F814W & $-$10.78 & 0.3  & $-$2.00 & 1.0 & 100\\
UGC 8508 & 1 & F606W,F814W & \nodata & 1.3  & $-$2.92 & 2.6 & 100\\
LGS3     & 1 & F555W,F814W & \nodata & \nodata & \nodata & 0.8 & 75\\
UGCA290  & 1 & F336W,F555W,F814W & \nodata & \nodata & \nodata & 6.7 & 100\\
M81Dwa   & 1 & F606W,F814W & \nodata & \nodata & \nodata & 3.5 & 100\\
\enddata
\tablecomments{This table lists each galaxy studied:
Column 2---the number of pointings observed with {\it HST}.
Column 3---the filters used.
Column 5---the diameter at which the B-band surface brightness of
the galaxy drops to 25 mag arcsec$^{-2}$. The M$_B$ and D$_{25}$
are taken from de Vaucouleurs et al.\ (1991).
Column 6---the
star formation rate, determined from the integrated H$\alpha$ luminosity,
normalized to the area within
D$_{25}$. Taken from Hunter \& Elmegreen (2002) and Hodge, Lee, \& Kennicutt
(1989).
Column 7---the distance to the galaxy.
Column 8---the percentage
of the D$_{25}$ area searched.}
\end{deluxetable}

\clearpage

\begin{deluxetable}{lcccccccc}
\tablewidth{0pt}
\tabletypesize{\footnotesize}
\tablecaption{Cluster Number Densities\label{tabcld}}
\tablehead{
\colhead{}
& \colhead{}
& \colhead{}
& \colhead{ND$_{-9.5}$\tablenotemark{b}}
& \colhead{}
& \colhead{ND$_{-10.5}$\tablenotemark{d}}
& \colhead{PA\tablenotemark{e}}
& \colhead{$i$\tablenotemark{e}}
& \colhead{} \\
\colhead{Galaxy}
& \colhead{N$_{cl}$}
& \colhead{N$_{-9.5}$\tablenotemark{a}}
& \colhead{(No.\ kpc$^{-2}$)}
& \colhead{N$_{-10.5}$\tablenotemark{c}}
& \colhead{(No.\ kpc$^{-2}$)}
& \colhead{(deg)}
& \colhead{(deg)}
& \colhead{E(B$-$V$_f$)\tablenotemark{f}}
}
\startdata
NGC 4214  & 29 &10--14 & 0.5--0.7 & 6--9 & 0.3--0.4 & 28 & 32    & 0.000 \\
NGC 2366  &  3 &  0    &  0       &  0   &    0     & 72 & 32    & 0.043 \\
DDO 50    &  3 &  0--1 &  0--0.1  & 0--1 & 0--0.1   & 52 & 20.5  & 0.023 \\
NGC 1705  & 16 &  7--8 & 1.9--2.2 &  4   &    1.1   & 43 & 49    & 0.045 \\
DDO 168   &  3 &  0--1 &  0--0.2  &  0   &    0     & 55 & $-$26 & 0.000 \\
DDO 165   &  4 &  0--4 &  0--1.3  & 0--1 & 0--0.3   & 60 & 91.5  & 0.010 \\
WLM       &  2 &  1    &  0.5     &  1   &    0.5   & 72 & $-$4  & 0.018 \\
Sextans A &  2 &  0--1 &  0--0.6  & 0--1 &  0--0.6  & 35 & 46    & 0.018 \\
IC 10\tablenotemark{g}    & 13 & 0 & 0 & 0 & 0 & \nodata & \nodata & \nodata \\
NGC 1569\tablenotemark{g} & 47 & 14 & 5.9 & 8 & 3.4 & \nodata & \nodata & \nodata \\
NGC 4449\tablenotemark{g} & 49 & 21 & 1.6 & 10 & 0.8 &\nodata & \nodata & \nodata \\
VIIZw403\tablenotemark{g} &  0 &  0 &  0  &  0 &  0  &\nodata & \nodata & \nodata \\
\enddata
\tablenotetext{a}{Number of populous and super-star clusters, taken to be
those with M$_V$ at 10 Myr brighter than $-9.5$. The range in numbers
represent the uncertainties in ages of some clusters.
The larger number is appropriate if an older age for some clusters is correct.
The lower limit in ranges is used in plots and discussion.}
\tablenotetext{b}{Number density of populous and super-star clusters
found within the D$_{25}$ area that was surveyed.}
\tablenotetext{c}{Number of super-star clusters, taken to be
those with M$_V$ at 10 Myr brighter than $-10.5$. The range in numbers
represent the uncertainties in ages of some clusters.
The larger number is appropriate if an older age for some clusters is correct.}
\tablenotetext{d}{Number density of populous and super-star clusters
found within the D$_{25}$ area that was surveyed.}
\tablenotetext{e}{The position angle and inclination of the
galaxy used to determine the distance of the cluster from the center
of the galaxy in the plane of the galaxy.}
\tablenotetext{f}{The foreground reddening to each galaxy from
Burstein \& Heiles (1984). To correct the cluster photometry
for reddening we assumed the total E(B$-$V) was 0.05 mag greater
than the foreground reddening.}
\tablenotetext{g}{These are taken from Lynds et al.\ (1998), Hunter et al.\ (2000),
Gelatt et al.\ (2001), and Hunter (2001).}
\end{deluxetable}

\clearpage

\begin{deluxetable}{lccccccc}
\tablewidth{0pt}
\tabletypesize{\footnotesize}
\tablecaption{Previously Known Nearby Super-Star Clusters for Comparison\label{tabssc}}
\tablehead{
\colhead{} & \colhead{R$_{\rm 0.5}$\tablenotemark{a}} & \colhead{}
& \colhead{Age} & \colhead{}
& \colhead{Mass} & \colhead{} & \colhead{} \\
\colhead{Cluster} & \colhead{(pc)} & \colhead{M$_V$}
& \colhead{(Myr)} & \colhead{M$_V$ at 10 Myr}
& \colhead{($10^4$ M\protect\solar)}
& \colhead{Galaxy M$_B$} & \colhead{Comment}
}
\startdata
NGC 1569-A & 2.3 & $-$14.0 & 5--7   & $-$12.8 & 30      & $-17.4$ & Starburst, interaction? \\
NGC 1569-B & 3.1 & $-$13.0 & 10--20 & $-$13.0 & \nodata & $-17.4$ & Starburst, interaction? \\
NGC 1705   & 3.4 & $-$14.0 & 7--20  & $-$14.0 & 8       & $-16.1$ & Starburst \\
NGC 4449-1 & 5.0 & $-$12.6 & 8--15  & $-$12.6 & \nodata & $-18.2$ & Interaction \\
LMC--R136  & 2.6 & $-$11.1 & 2      & $-$10.5 & 6       & $-18.1$ & Interaction, bar \\
WLM\tablenotemark{b} &  3: & $-$8.9  & 14800  & $-$14.2 & \nodata & $-14.2$ & \nodata \\
NGC 2366?  & Embedded & \nodata & $<1$ & \nodata & \nodata & $-16.5$ & \nodata \\
NGC 5253?  & Embedded & \nodata & $<1$ & \nodata & \nodata & $-17.7$ & Post-starburst \\
He2-10?  & Embedded & \nodata & $<1$ & \nodata & \nodata & $-18.1$ & Interaction \\
NGC 4214?  & Embedded & \nodata & 4--5 & \nodata & \nodata & $-18.4$ & \nodata \\
MW globulars\tablenotemark{c}& 1--8 & -7.4 & 10000 & $-$12.9 & 10--100 & \nodata & \nodata \\
\enddata
\tablenotetext{a}{The cluster half-light radius.}
\tablenotetext{b}{Age taken from Hodge et al.\ (1999). Radius is half of the
FWHM rather than R$_{0.5}$.}
\tablenotetext{c}{These are typical values for globular clusters in the Milky Way.}
\tablecomments{References---
He2-10: Conti \& Vacca 1994; Kobulnicky \& Johnson 1999;
LMC: Hunter et al.\ 1995;
WLM: Hodge et al.\ 1999;
NGC 1569: O'Connell, Gallagher, \& Hunter 1994;
Ho \& Filippenko 1996a; Hunter et al.\ 2000;
NGC 1705: Meurer et al.\ 1992; O'Connell et al.\ 1994;
Ho \& Filippenko 1996b;
NGC 2366: Drissen et al.\ 2000;
NGC 4214: Leitherer et al.\ 1996;
NGC 4449: Gelatt, Hunter, \& Gallagher 2001;
NGC 5253: Turner, Ho, \& Beck 1998;
Milky Way globular clusters:
van den Bergh et al. 1991, Harris 1991.
}
\end{deluxetable}

\clearpage

\begin{deluxetable}{ccccccccccc}
\tablewidth{0pt}
\tabletypesize{\scriptsize}
\tablecaption{NGC 4214 Cluster Parameters\label{tabn4214}}
\tablehead{
\colhead{} & \colhead{} & \colhead{}
& \colhead{}
& \colhead{}
& \colhead{} & \colhead{}
& \colhead{R\tablenotemark{b}} & \colhead{Age\tablenotemark{c}}
& \colhead{}
& \colhead{FWHM)} \\
\colhead{Cluster} & \colhead{m$_{F555,0}$} & \colhead{M$_V$\tablenotemark{a}}
& \colhead{(V$-$I)$_0$\tablenotemark{a}}
& \colhead{(U$-$V)$_0$\tablenotemark{a}}
& \colhead{RA (2000)} & \colhead{Dec (2000)} &
\colhead{(kpc)} & \colhead{(Myr)}
& \colhead{M$_{\rm V}$ at 10 Myr}
& \colhead{(pc)}
}
\startdata
1 & 20.28 & $-$8.33  & 0.77  & 0.84  & 12:15:29.83 &      36:20:05.0   & 2.9  & 1000 & $-$11.9 & 4.8\\
 & & (0.01) & (0.02) & (0.18) &&&&&&\\
2 & 20.30 & $-$8.29  & 0.80  & 0.65  & 12:15:31.20 &      36:19:37.2   & 2.4  & 1000 & $-$11.8 & 5.5\\
 & & (0.01) & (0.02) & (0.10) &&&&&&\\
3 & 21.63 & $-$6.96  & 0.87  & \nodata & 12:15:34.23 &     36:19:13.6   & 1.5  & 9$-$14/1000 & $-$6.8/$-$10.5 & 5.7\\
 & & (0.03) & (0.04) & \nodata &&&&&&\\
4 & 18.88 & $-$9.71  & 0.95  & 0.64  & 12:15:34.47 &      36:20:19.0   & 1.7  & 1000 & $-$13.3 & 5.3\\
 & & (0.01) & (0.01) & (0.04) &&&&&&\\
5 & 20.32 & $-$8.20 & $-$0.75  & $-$0.83  & 12:15:34.58 &         36:20:00.9   & 1.5  & 5: & \nodata & 5.9\\
 & & (0.02) & (0.05) & (0.04) &&&&&&\\
6 & 21.60 & $-$6.98  & 0.24  & \nodata & 12:15:34.71 &      36:19:10.9   & 1.4  & 7$-$19 & $-$5.7/$-$7.4&5.9\\
 & & (0.03) & (0.06) & \nodata &&&&&&\\
7 & 21.51 & $-$7.07  & 1.21  & \nodata & 12:15:35.06 &     36:18:31.0   & 1.9  & 12$-$14 & $-$7.6 & 5.1\\
 & & (0.03) & (0.04) & \nodata &&&&&&\\
8 & 21.05 & $-$7.51  & 1.70  & \nodata & 12:15:35.22 &      36:18:39.5   & 2.1  & \nodata & \nodata & 6.4\\
 & & (0.02) & (0.03) & \nodata &&&&&&\\
9 & 20.17 & $-$8.40  & 0.36  & \nodata  & 12:15:36.25 &      36:19:59.5 &   1.0  & 7$-$22 &$-$7.2/$-$9.1&9.5\\
 & & (0.01) & (0.02) & \nodata &&&&&&\\
10 & 17.25 & $-$11.33 & 0.49  & 0.05  & 12:15:37.22 &     36:19:56.4  &  0.7  & 220 &$-$13.7 & 8.4\\
 & & (0.01) & (0.01) & (0.01) &&&&&&\\
11 & 20.76 & $-$7.82  & 0.36  & 0.08  & 12:15:37.27 &     36:18:54.0  &  1.3  & 7/250 &$-$6.6/$-$10.3 & 5.9\\
 & & (0.01) & (0.01) & (0.04) &&&&&&\\
12 & 20.91 & $-$7.61  & $-$0.63 &  $-$0.31 & 12:15:37.60 &        36:19:01.0  &  0.9  & 6: & \nodata & 4.8\\
 & & (0.01) & (0.00) & (0.03) &&&&&&\\
13 & 16.54 & $-$12.04 & 0.87  & 0.48  & 12:15:38.16 &     36:19:44.4  &  0.3  & 700--1000 &$-$15.4 & 5.1\\
 & & (0.00) & (0.01) & (0.01) &&&&&&\\
14 & 20.84 & $-$7.73  & 1.64  & \nodata & 12:15:38.51 &     36:21:08.6   & 2.2  & \nodata & \nodata & 6.7\\
 & & (0.02) & (0.03) & \nodata &&&&&&\\
15 & 21.11 & $-$7.47  & 0.46  & 0.11  & 12:15:39.63 &     36:18:19.0 &   2.3  & 270 &$-$10.0 & 9.4\\
 & & (0.01) & (0.02) & (0.05) &&&&&&\\
16 & 18.66 & $-$9.92  & 1.08  & 0.37  & 12:15:40.15 &     36:19:38.0  &  0.3  & 10 & $-$9.9 & 8.9\\
 & & (0.01) & (0.01) & (0.03) &&&&&&\\
17 & 21.05 & $-$7.53  & 0.87  & 0.97  &  12:15:40.67 &    36:20:02.1  &  0.8  & 1000 & $-$11.1 & 9.6\\
 & & (0.02) & (0.03) & (0.10) &&&&&&\\
18 & 19.77 & $-$8.76 & $-$0.55  & $-$0.49  & 12:15:41.06 &        36:19:29.7 &   0.6  & 6 & $-$7.5 & 5.8\\
 & & (0.02) & (0.03) & (0.03) &&&&&&\\
19 & 19.79 & $-$8.78   & 0.22 & 0.97  &  12:15:41.27 &   36:19:34.3 &   0.7  & \nodata & \nodata & 6.4\\
 & & (0.01) & (0.01) & (0.05) &&&&&&\\
20 & 19.97 & $-$8.61  & 0.76  & 1.02  & 12:15:41.35 &     36:19:02.8  &  1.1  & 8 & $-$7.9 & 10.5 \\
 & & (0.01) & (0.02) & (0.06) &&&&&&\\
21 & 20.98 & $-$7.60  & 0.81  & \nodata & 12:15:41.39 &    36:19:38.2   & 0.7  & 8$-$15/1000 & $-$7.4/$-$11.2 & 8.1\\
 & & (0.03) & (0.05) & \nodata &&&&&&\\
22 & 21.46 & $-$7.12   & 0.27  & $-$0.05 & 12:15:41.96 &  36:19:33.1  &  0.9  & 7 & $-$5.9 & 5.2\\
 & & (0.03) & (0.06) & (0.12) &&&&&&\\
23 & 21.47 & $-$7.10   & 0.24  & \nodata &  12:15:42.63 &   36:19:34.8   & 1.1  & 7$-$19 &$-$5.8/$-$7.5&6.2\\
 & & (0.04) & (0.06) & \nodata &&&&&&\\
24 & 20.53 & $-$8.05  &  0.26  & 0.13 & 12:15:43.37 &     36:18:49.1  &  1.9  & 7/300 & $-$6.8/$-$10.7 & 5.7\\
 & & (0.02) & (0.03) & (0.08) &&&&&&\\
25 & 20.51 & $-$8.07   & 0.28  & $-$0.00 & 12:15:43.39 &  36:18:48.8   & 1.9  & 7 & $-$6.8 & 6.0\\
 & & (0.01) & (0.01) & (0.03) &&&&&&\\
26 & 21.32 & $-$7.26   & 0.40  & 0.39 &  12:15:43.42 &    36:18:42.0   & 1.8  & 600 & $-$10.4 & 8.5\\
 & & (0.02) & (0.03) & (0.08) &&&&&&\\
27 & 21.27 & $-$7.31   & 0.45  & 0.16 &  12:15:44.73 &    36:19:04.8   & 1.8  & 300 & $-$9.9 & 4.9\\
 & & (0.02) & (0.04) & (0.13) &&&&&&\\
28 & 19.12 & $-$9.40  & $-$0.71  & $-$0.37  &  12:15:47.70 &      36:19:00.8   & 2.7  & 6: & \nodata & 9.3\\
 & & (0.01) & (0.02) & (0.03) &&&&&&\\
29 & 19.86 & $-$8.72   & 0.37  & 0.76  & 12:15:47.95 &    36:19:01.7   & 2.8  & \nodata & \nodata & 9.2\\
 & & (0.01) & (0.02) & (0.08) &&&&&&\\
\enddata
\tablenotetext{a}{The uncertainties for each photometric measurement are
in the row immediately following it.}
\tablenotetext{b}{R is the radius of the cluster from the center of the
galaxy in the plane of the galaxy.}
\tablenotetext{c}{Values followed with a ``:'' indicate clusters whose
photometry placed them far from the cluster evolutionary tracks.}
\end{deluxetable}

\clearpage

\begin{deluxetable}{cccccccccc}
\tablewidth{0pt}
\tabletypesize{\footnotesize}
\tablecaption{NGC 2366 Cluster Parameters\label{tabn2366}}
\tablehead{
\colhead{} & \colhead{}
& \colhead{}
& \colhead{}
& \colhead{} & \colhead{}
& \colhead{R\tablenotemark{c}} & \colhead{Age\tablenotemark{d}}
& \colhead{}
& \colhead{FWHM)} \\
\colhead{Cluster\tablenotemark{a}} & \colhead{m$_{F555,0}$} & \colhead{M$_V$\tablenotemark{b}}
& \colhead{(B$-$V)$_0$\tablenotemark{b}}
& \colhead{RA (2000)} & \colhead{Dec (2000)} &
\colhead{(kpc)} & \colhead{(Myr)}
& \colhead{M$_{\rm V}$ at 10 Myr}
& \colhead{(pc)}
}
\startdata
B & 18.21 &$-$9.51 &$-$0.26 &7:28:43.47 & 69:11:22.7 & 1.2 & 3 & $-$8.8 & 4.6\\
  &     & (0.01) & (0.05) &&&&&&\\
1 & 19.15 &$-$8.52 &$-$0.26 &7:28:43.23 & 69:11:21.7 & 1.2 & 3 & $-$7.8 & 4.6\\
  &     & (0.01) & (0.05) &&&&&&\\
2 & 19.11 &$-$8.47 & 0.68 &7:28:54.52 & 69:11:12.2 & 1.4 & \nodata & \nodata & 14.2\\
  &     & (0.01) & (0.05) &&&&&&\\
\enddata
\tablenotetext{a}{Cluster B is as identified by Drissen et al.\ (2000). The age of this
cluster is from them. They refer to our cluster 1 as a star, but the object is clearly
resolved with respect to a PSF in the image and meets our
criteria for a cluster.}
\tablenotetext{b}{The uncertainties for each photometric measurement are
in the row immediately following it.}
\tablenotetext{c}{R is the radius of the cluster from the center of the
galaxy in the plane of the galaxy.}
\tablenotetext{d}{Values followed with a ``:'' indicate clusters whose
photometry placed them far from the cluster evolutionary tracks.}
\end{deluxetable}

\clearpage

\begin{deluxetable}{cccccccccc}
\tablewidth{0pt}
\tabletypesize{\footnotesize}
\tablecaption{DDO 50 Cluster Parameters\label{tabd50}}
\tablehead{
\colhead{} & \colhead{}
& \colhead{}
& \colhead{}
& \colhead{} & \colhead{}
& \colhead{R\tablenotemark{b}} & \colhead{Age\tablenotemark{c}}
& \colhead{}
& \colhead{FWHM)} \\
\colhead{Cluster} & \colhead{m$_{F606,0}$} & \colhead{M$_R$\tablenotemark{a}}
& \colhead{(R$-$I)$_0$\tablenotemark{a}}
& \colhead{RA (2000)} & \colhead{Dec (2000)} &
\colhead{(kpc)} & \colhead{(Myr)}
& \colhead{M$_{\rm V}$ at 10 Myr}
& \colhead{(pc)}
}
\startdata
1  & 21.24 & $-$6.89 & 0.81 & 8:18:58.27 &  70:45:06.6 & 5.8   & \nodata & \nodata & 2.1 \\
   &        & (0.02) & (0.05) &  &  &  &  && \\
2  & 19.84 & $-$8.11  & 0.73 & 8:19:00.75 &  70:44:23.3 & 6.2   & \nodata & \nodata & 4.6\\
   &        & (0.01) & (0.05) &  &  &  &  && \\
3  & 20.57 & $-$7.17  & 0.37 & 8:18:57.99 &  70:44:06.1 & 6.5   & $>$80 & $-$8.5/$-$10.5 & 9.0\\
   &        & (0.01) & (0.06) &  &  &  &  &&\\
\enddata
\tablenotetext{a}{The uncertainties for each photometric measurement are
in the row immediately following it.}
\tablenotetext{b}{R is the radius of the cluster from the center of the
galaxy in the plane of the galaxy.}
\tablenotetext{c}{Values followed with a ``:'' indicate clusters whose
photometry placed them far from the cluster evolutionary tracks.}
\end{deluxetable}

\clearpage

\begin{deluxetable}{ccccccccccc}
\tablewidth{0pt}
\tabletypesize{\scriptsize}
\tablecaption{NGC 1705 Cluster Parameters\label{tabn1705}}
\tablehead{
\colhead{} & \colhead{} & \colhead{}
& \colhead{}
& \colhead{}
& \colhead{} & \colhead{}
& \colhead{R\tablenotemark{b}} & \colhead{Age\tablenotemark{c}}
& \colhead{}
& \colhead{FWHM)} \\
\colhead{Cluster} & \colhead{m$_{F555,0}$} & \colhead{M$_V$\tablenotemark{a}}
& \colhead{(V$-$I)$_0$\tablenotemark{a}}
& \colhead{(B$-$V)$_0$\tablenotemark{a}}
& \colhead{RA (2000)} & \colhead{Dec (2000)}
& \colhead{(kpc)} & \colhead{(Myr)}
& \colhead{M$_{\rm V}$ at 10 Myr}
& \colhead{(pc)}
}
\startdata
A\tablenotemark{d} & \nodata    &$-$13.71&  0.96 & \nodata & 4:54:13.24 & $-$53:21:36.1 & 0.2 & 7$-$20 &$-$14 & \nodata \\
 & & \nodata & \nodata & \nodata &&&&&&\\
%1 & 22.57 &$-$6.24&  1.11& 0.81&    4:54:05.70&   $-$53:20:42.6&  2.2&       1000 &    $-$9.8 & 10.8\\
%& & (0.01) & (0.01) & (0.03) &&&&&&\\
%2 & 22.59 &$-$6.22&  0.70& 0.73&    4:54:06.07&   $-$53:21:27.9&  2.0&       1000 & $-$9.8 & 12.9\\
% & & (0.01) & (0.01) & (0.04) &&&&&&\\
%3 & 22.10 &$-$6.68& $-$0.20& $-$0.23&    4:54:06.43&   $-$53:21:39.1&  2.0&       4 & $-$5.8 & 5.1 \\
% & & (0.01) & (0.01) & (0.01) &&&&&&\\
1 & 22.34 &$-$6.40&  2.4& 1.5&    4:54:08.62&   $-$53:21:22.7&  1.3&       \nodata & \nodata & 5.0 \\
 & & (0.01) & (0.01) & (0.06) &&&&&&\\
%5 & 22.41 &$-$6.40&  1.13& 0.58&    4:54:10.13&   $-$53:20:20.3&  2.1&       1000      & $-$10.0 & 5.3\\
% & & (0.01) & (0.01) & (0.02) &&&&&&\\
%6 & 22.61 &$-$6.20&  1.12& 0.61&    4:54:10.39&   $-$53:20:19.4&  2.1&       1000     & $-$9.8 & 6.5\\
% & & (0.01) & (0.01) & (0.03) &&&&&&\\
2 & 22.04 &$-$6.77&  0.94& 0.72&    4:54:10.89&   $-$53:21:59.2&  1.3&       1000      & $-$10.3 & 11.3\\
 & & (0.01) & (0.02) & (0.03) &&&&&&\\
3 & 21.46 &$-$7.35&  0.89& 0.67&    4:54:11.95&   $-$53:21:49.9&  0.8&       1000      & $-$10.9 & 7.1\\
 & & (0.01) & (0.02) & (0.03) &&&&&&\\
4 & 22.50 &$-$6.31&  0.76& \nodata &      4:54:11.97&   $-$53:21:50.0&  0.6&  8$-$15/1000 & $-$6/$-$9.9 & 5.4\\
 & & (0.01) & (0.02) & \nodata &&&&&&\\
5& 22.38 &$-$6.43&  0.90& 0.74&    4:54:12.54&   $-$53:21:52.5&  0.6&       1000 & $-$10.0 & 4.8\\
 & & (0.01) & (0.01) & (0.02) &&&&&&\\
6& 20.95 &$-$7.87&  0.84& 0.63&    4:54:12.80&   $-$53:21:18.3&  0.6&       1000 & $-$11.4 & 5.8\\
 & & (0.01) & (0.01) & (0.02) &&&&&&\\
7& 20.90 &$-$7.91&  0.83& 0.66&    4:54:12.94&   $-$53:21:45.2&  0.4&       1000 & $-$11.5 & 5.5\\
 & & (0.01) & (0.01) & (0.01) &&&&&&\\
8& 22.77 &$-$6.04&  0.55& 0.20&    4:54:13.11&   $-$53:21:51.3&  0.5&       15/150 & $-$5.9/$-$8.1 & 3.9\\
 & & (0.02) & (0.02) & (0.02) &&&&&&\\
9& 22.07 &$-$6.74&  0.85& 0.63&    4:54:13.28&   $-$53:21:48.4&  0.5&       1000 & $-$10.3 & 4.4\\
 & & (0.01) & (0.02) & (0.03) &&&&&&\\
10& 20.98 &$-$7.83&  0.30& 0.10&    4:54:13.68&   $-$53:21:48.0&  0.4&       7 & $-$6.7 & 5.4\\
 & & (0.01) & (0.02) & (0.02) &&&&&&\\
11& 21.20 &$-$7.60&  0.24& 0.15&    4:54:13.95&   $-$53:21:43.9&  0.3&       7 & $-$6.3 & 4.5\\
 & & (0.01) & (0.02) & (0.02) &&&&&&\\
12& 21.10 &$-$7.71&  0.26& 0.10&    4:54:14.35&   $-$53:21:32.7&  0.1&       7 & $-$6.5 & 5.9\\
 & & (0.07) & (0.01) & (0.01) &&&&&&\\
13& 21.61 &$-$7.20&  0.47& 0.13&    4:54:14.47&   $-$53:21:43.3&  0.4&       15 & $-$7.1 & 5.0\\
 & & (0.01) & (0.02) & (0.02) &&&&&&\\
14& 21.59 &$-$7.22&  0.30& 0.12&    4:54:14.60&   $-$53:21:43.5&  0.4&       7 & $-$6.1 & 4.8\\
 & & (0.01) & (0.02) & (0.02) &&&&&&\\
15& 22.29 &$-$6.50&  1.65& 1.71&    4:54:14.74&   $-$53:22:17.8&  1.3&       \nodata & \nodata & 5.1\\
 & & (0.01) & (0.01) & (0.05) &&&&&&\\

\enddata
\tablenotetext{a}{The uncertainties for each photometric measurement are
in the row immediately following it.}
\tablenotetext{b}{R is the radius of the cluster from the center of the
galaxy in the plane of the galaxy.}
\tablenotetext{c}{Values followed with a ``:'' indicate clusters whose
photometry placed them far from the cluster evolutionary tracks.}
\tablenotetext{d}{The data for this cluster are taken from O'Connell et al.\ (1994).}
\end{deluxetable}

\clearpage

\begin{deluxetable}{cccccccccc}
\tablewidth{0pt}
\tabletypesize{\footnotesize}
\tablecaption{DDO 168 Cluster Parameters\label{tabd168}}
\tablehead{
\colhead{} & \colhead{}
& \colhead{}
& \colhead{}
& \colhead{} & \colhead{}
& \colhead{R\tablenotemark{b}} & \colhead{Age\tablenotemark{c}}
& \colhead{}
& \colhead{FWHM)} \\
\colhead{Cluster} & \colhead{m$_{F606,0}$} & \colhead{M$_R$\tablenotemark{a}}
& \colhead{(R$-$I)$_0$\tablenotemark{a}}
& \colhead{RA (2000)} & \colhead{Dec (2000)} &
\colhead{(kpc)} & \colhead{(Myr)}
& \colhead{M$_{\rm V}$ at 10 Myr}
& \colhead{(pc)}
}
\startdata
1 & 20.88 &$-$7.06  &0.37 &13:14:26.01 & 45:54:52.7 & 1.1 & 8$-$45/$>$530 & $-$6.0/$-$10.1&7.4\\
  &      & (0.01) & (0.05) & & & & & &\\
2 & 21.00 &$-$7.13  &0.71 &13:14:28.54 & 45:54:36.0 & 1.3 & 12 & $-$6.9 & 3.6\\
  &      & (0.01) & (0.05) & & & & & &\\
3 & 20.30 & $-$7.78 & 0.85 & 13:14:30.90 & 45:54:59.7 & 1.2 & 10 & $-$7.3 & 4.3 \\
  &      & (0.01) & (0.02) & & & & & &\\
\enddata
\tablenotetext{a}{The uncertainties for each photometric measurement are
in the row immediately following it.}
\tablenotetext{b}{R is the radius of the cluster from the center of the
galaxy in the plane of the galaxy.}
\tablenotetext{c}{Values followed with a ``:'' indicate clusters whose
photometry placed them far from the cluster evolutionary tracks.}
\end{deluxetable}

\clearpage

%\begin{deluxetable}{cccccccccc}
%\tablewidth{0pt}
%\tablecaption{IC 1613 Cluster Parameters\label{tabic1613}}
%\tablehead{
%\colhead{} & \colhead{F555$_0$} & \colhead{M$_V$\tablenotemark{a}}
%& \colhead{V$-$I\tablenotemark{a}}
%& \colhead{RA} & \colhead{Dec} & \colhead{R\tablenotemark{b}} & \colhead{Age in Myr}
%& \colhead{}
%& \colhead{FWHM (pc)}
%}
%\startdata
%1 & 18.19 &$-$6.23 & 0.79 &1:04:29.86 &2:03:05.4  & 1.4 & 90: &...&1.3\\
%  &      & (0.00) & (0.00) &&&&&&\\
%\enddata
%\tablenotetext{a}{The uncertainties for each photometric measurement are
%in the row immediately following it.}
%\tablenotetext{b}{R is the radius of the cluster from the center of the
%galaxy in the plane of the galaxy.}
%\tablenotetext{c}{Values followed with a ``:'' indicate clusters whose
%photometry placed them far from the cluster evolutionary tracks.}
%\end{deluxetable}
%
%\clearpage

\begin{deluxetable}{cccccccccc}
\tablewidth{0pt}
\tabletypesize{\footnotesize}
\tablecaption{DDO 165 Cluster Parameters\label{tabd165}}
\tablehead{
\colhead{} & \colhead{}
& \colhead{}
& \colhead{}
& \colhead{} & \colhead{}
& \colhead{R\tablenotemark{b}} & \colhead{Age\tablenotemark{c}}
& \colhead{}
& \colhead{FWHM)} \\
\colhead{Cluster} & \colhead{m$_{F606,0}$} & \colhead{M$_R$\tablenotemark{a}}
& \colhead{(R$-$I)$_0$\tablenotemark{a}}
& \colhead{RA (2000)} & \colhead{Dec (2000)} &
\colhead{(kpc)} & \colhead{(Myr)}
& \colhead{M$_{\rm V}$ at 10 Myr}
& \colhead{(pc)}
}
\startdata
1 & 19.17 &$-$8.14 & 0.39 & 13:06:29.33 & 67:42:27.7 &0.3 &8$-$15/$>$930 &$-$7.1/$-$11.3&6.1\\
  &        & (0.01) & (0.05) & & & & & &\\
2 & 18.76 &$-$8.54 & 0.38 & 13:06:31.00 & 67:42:17.8 &0.5 &8$-$15 $>$670 &$-$7.5/$-$11.8&2.8\\
  &        & (0.00) & (0.05) & & & & & &\\
3 & 20.69 &$-$6.54 & 0.25 & 13:06:30.40 & 67:41:52.6 &1.0 &7$-$720 &$-$5.2/$-$9.5&3.6\\
  &        & (0.01) & (0.05) & & & & & &\\
4 & 19.94 &$-$7.24 & 0.18 & 13:06:22.75 & 67:42:02.9 &0.7 &7$-$19 &$-$5.9/$-$7.5&3.1\\
  &        & (0.01) & (0.05) & & & & & &\\
\enddata
\tablenotetext{a}{The uncertainties for each photometric measurement are
in the row immediately following it.}
\tablenotetext{b}{R is the radius of the cluster from the center of the
galaxy in the plane of the galaxy.}
\tablenotetext{c}{Values followed with a ``:'' indicate clusters whose
photometry placed them far from the cluster evolutionary tracks.}
\end{deluxetable}

\clearpage

\begin{deluxetable}{cccccccccc}
\tablewidth{0pt}
\tabletypesize{\footnotesize}
\tablecaption{WLM Cluster Parameters\label{tabwlm}}
\tablehead{
\colhead{} & \colhead{}
& \colhead{}
& \colhead{}
& \colhead{} & \colhead{}
& \colhead{R\tablenotemark{b}} & \colhead{Age\tablenotemark{c}}
& \colhead{}
& \colhead{FWHM)} \\
\colhead{Cluster} & \colhead{m$_{F555,0}$} & \colhead{M$_V$\tablenotemark{a}}
& \colhead{(V$-$I)$_0$\tablenotemark{a}}
& \colhead{RA (2000)} & \colhead{Dec (2000)}
& \colhead{(kpc)} & \colhead{(Myr)}
& \colhead{M$_{\rm V}$ at 10 Myr}
& \colhead{(pc)}
}
\startdata
1\tablenotemark{d} & 16.36 &$-$8.88 & 0.81 & 0:01:49.48 & $-$15:27:30.7 & 3.0 &14800 & $-$14.4 &5.6 \\
  &     & (0.00) & (0.00) &&&&&&\\
2 & 18.95 &$-$6.29 & 0.86 & 0:01:55.50 & $-$15:24:43.2 & 1.1 & 1000: & \nodata &1.2\\
  &     & (0.00) & (0.00) &&&&&&\\
\enddata
\tablenotetext{a}{The uncertainties for each photometric measurement are
in the row immediately following it.}
\tablenotetext{b}{R is the radius of the cluster from the center of the
galaxy in the plane of the galaxy.}
\tablenotetext{c}{Values followed with a ``:'' indicate clusters whose
photometry placed them far from the cluster evolutionary tracks.}
\tablenotetext{d}{This cluster is a known globular cluster in WLM. The age is
taken from Hodge et al. (1999).}
\end{deluxetable}

\clearpage

\begin{deluxetable}{ccccccccccc}
\tablewidth{0pt}
\tabletypesize{\scriptsize}
\tablecaption{Sextans A Cluster Parameters\label{tabsexa}}
\tablehead{
\colhead{} & \colhead{} & \colhead{}
& \colhead{}
& \colhead{}
& \colhead{} & \colhead{}
& \colhead{R\tablenotemark{b}} & \colhead{Age\tablenotemark{c}}
& \colhead{}
& \colhead{FWHM)} \\
\colhead{Cluster} & \colhead{m$_{F555,0}$} & \colhead{M$_V$\tablenotemark{a}}
& \colhead{(V$-$I)$_0$\tablenotemark{a}}
& \colhead{(B$-$V)$_0$\tablenotemark{a}}
& \colhead{RA (2000)} & \colhead{Dec (2000)} &
\colhead{(kpc)} & \colhead{(Myr)}
& \colhead{M$_{\rm V}$ at 10 Myr}
& \colhead{(pc)}
}
\startdata
1 & 19.00 &$-$7.12 & 0.68 & \nodata  &10:10:51.81   &$-$4:40:57.8  &1.2 & $>$75 & $-$8.7/$-$10.9 & 1.6\\
  &     & (0.00) & (0.00) & \nodata &&&&&&\\
%2 & 18.39 &$-$7.73 & 0.84 &0.37 &10:11:01.85 &$-$4:42:33.1  &0.4 & 1000 & $-$11.5 & 1.6\\
%  &     & (0.00) & (0.00) & (0.01) &&&&&&\\
2 & 19.50 &$-$6.58 &$-$0.42 &$-$0.30 &10:11:05.27  & $-$4:42:40.5  &0.8 & 3: & \nodata & 1.6\\
  &     & (0.00) & (0.01) & (0.01) &&&&&&\\
%4 & 18.53 &$-$7.58 & 1.37 &1.58 &10:11:05.44  & $-$4:41:57.2  &0.6& 1000: & \nodata  & 1.5\\
%  &     & (0.00) & (0.00) & (0.01) &&&&&&\\
\enddata
\tablenotetext{a}{The uncertainties for each photometric measurement are
in the row immediately following it.}
\tablenotetext{b}{R is the radius of the cluster from the center of the
galaxy in the plane of the galaxy.}
\tablenotetext{c}{Values followed with a ``:'' indicate clusters whose
photometry placed them far from the cluster evolutionary tracks.}
\end{deluxetable}

\end{document}